\documentclass[twocolumn,amsmath,superscriptaddress,amssymb,nopacs,nofootinbib,prb]{revtex4-2}

\usepackage[usenames,dvips]{color}
\usepackage{graphicx}
\usepackage{dcolumn}
\usepackage{bm}
\usepackage{mathtools}
\usepackage{epstopdf}
\usepackage{esint}
\usepackage{xcolor}
\usepackage{dsfont}
\usepackage{nicefrac}
\usepackage{tikz}
\usepackage{enumitem}
\usepackage[normalem]{ulem}

\usepackage[colorlinks=true,citecolor=blue,linkcolor=magenta, urlcolor=magenta]{hyperref}

\newcommand{\bea}{\begin{eqnarray}}
\newcommand{\eea}{\end{eqnarray}}
\newcommand{\bt}{\textbf}

\newcommand{\ph}{\phantom{.}}

\newcommand{\noi}{\noindent}
\newcommand{\no}{\nonumber}

\begin{document}
\def\v#1{{\bf #1}}

\title{Theory of Berry Singularity Markers:\\Diagnosing Topological Phase Transitions via Lock-In Tomography}

\author{Panagiotis Kotetes}\email{kotetes@baqis.ac.cn}
\affiliation{Beijing Academy of Quantum Information Sciences, Beijing 100193, China}

\affiliation{CAS Key Laboratory of Theoretical Physics, Institute of Theoretical Physics, Chinese Academy of Sciences, Beijing 100190, China}

\vskip 1cm
\begin{abstract}
This work brings forward an alternative approach for inferring the topological character of phase transitions in insulators. The method relies on subjecting the system of interest to a set of external fields which consist of two parts, i.e., a weak spatiotemporally slowly-varying component on top of a constant offset. This process maps the Berry singularities dictating the bulk band structure of the insulator to a synthetic space spanned by the control parameters of the external fields. By measuring the response of the system to the weak part of the perturbations, when these are additionally chosen to form spacetime textures, one can construct a quantity that is here-termed Berry singularity marker (BSM). The BSM enables the Berry singularity detection as it becomes nonzero only in the close vicinity of a Berry singularity and is equal to its charge. Moreover, near the Berry singularities, the BSM encodes the quantum metric tensor of the system. An advantage of the present method is that it can be implemented in a lock-in fashion, thus allowing to ``filter out''  disorder effects, therefore opening the door to the resilient diagnosis of topological phase transitions. \end{abstract}

\maketitle

\section{Introduction}\label{sec:SectionI}

The discovery of new strategies to unambiguously detect the non\-tri\-vial character of topological materials and hybrid devices remains a pressing challenge in this timely research field. For instance, recent experiments~\cite{FrolovUbiquitous,YuQuasiMZM,Katsaros,HaoZhang} and theo\-re\-ti\-cal stu\-dies~\cite{CXLiu,HellInterferometry,Moore,CXLiu2,Reeg,Woods,Vuik,PanDasSarma,PradaRev,Hess} have highlighted the difficulties that are encountered when trying to distinguish Majorana zero modes (MZMs) from topologically-trivial Andreev bound states. While the latter originate from smooth confinement potentials or di\-sorder~\cite{Brouwer,Akhmerov,JLiu,Bagrets,Prada2012,DimaPeak,RoyTewari,SauSarma,PinningInteractions}, they may nonetheless become pinned near zero energy for an extended range in pa\-ra\-meter space. Notably, such hurdles arise for the majority of topological sy\-stems, since the bulk nontrivial to\-po\-lo\-gy is ty\-pi\-cal\-ly inferred in experiments indirectly, that is, by pro\-bing the to\-po\-lo\-gi\-cal\-ly-protected modes which emerge at boun\-daries where systems with different topological properties meet~\cite{Altland,SchnyderClassi,KitaevClassi,Ryu,TeoKane}. While this pursuit is certainly ju\-sti\-fied by the desire to harness the exotic pro\-per\-ties of the arising modes for cutting edge applications~\cite{HasanKane,QiZhang}, this approach has una\-voi\-da\-bly hindered the de\-ve\-lop\-ment of possible me\-thods to detect the bulk topology directly, i.e., by em\-ploying bulk-only measurements without involving the to\-po\-lo\-gi\-cal\-ly-protected modes.

Recently, however, a number of theoretical proposals have appeared that advocate for alternative ways to infer the bulk topological properties of the sy\-stem of interest. For instance, in connection to MZMs, it has been proposed to probe the nonlocal conductance~\cite{Roshdal,Jeroen,Menard3Terminal,3Terminal,PikulinProtocol,Seridonio}, the shot noise~\cite{Demler,Strubi,EggerNoise,Manousakis,BeenakkerNoise,SimonShotNoise}, or, the equili\-brium spin-susceptibility~\cite{Pakizer}. The spin-orbital magnetic suscep\-ti\-bi\-li\-ty has been also proposed as a bulk-topology detection means for topological insulators~\cite{Ogata}. Another proposed experimental method that has attracted attention is to infer properties of the bulk topology by studying the bound states of the system~\cite{Slager,Metamorphosis,SemiAnalytical} which are induced by impurities~\cite{BalatskyRMP}. The last development that I wish to mention here concerns the mapping out of the quantum geo\-metry~\cite{Provost,Marzari,Souza} which holds information for the quantum critical and topological pro\-per\-ties of the bulk system~\cite{CPSun,Zanardi,HQLin,Ma1,Matsuura,Ma2,Montambaux}. For an insulator, its ``ima\-ging'' can be achieved by inducing interband transitions to the system~\cite{Neupert,KolodrubetzPRB,Polkovnikov,Ozawa,Cooper,Gavensky,Rastelli,Klees,WeiChen}. To date, the quantum geometry appears to have been successfully measured in various solid-state devices~\cite{Martinis,Schroer,Weitenberg,BerryTomography,qutrit,QuantumMetricSCqubit}.

In this Article, I propose an alternative approach which aims at directly experimentally diagnosing the bulk to\-po\-lo\-gy of a system using bulk-only equilibrium or quasi-equilibrium measurements. The present method relies on the observation that a phase transition se\-pa\-ra\-ting two topologically-distinct phases is also endowed with topologically-nontrivial properties itself. Such an aspect appears to have been missed in the abovementioned works. Specifically, the target of the here-proposed diag\-nostic protocol is to map out the topological pro\-per\-ties of the Berry singularities which characterize the bulk Hamiltonian and me\-dia\-te the va\-rious to\-po\-lo\-\-gi\-cal phase transitions. The read-out of the to\-po\-lo\-gi\-cal charges and locations of Berry singularities is achieved by suitable mappings to a synthetic space which is spanned by a set of $N$ experimentally-controllable quantities compri\-sing a vector $\bm{F}=(F_1,F_2,\dots, F_N)$. $\bm{F}$ is assumed to take the form $\bm{F}(\bm{X})=\bm{B}+\bm{{\cal B}}(\bm{X})$ where $\bm{B}$ is a constant offset, while $\bm{{\cal B}}(\bm{X})$ exhibits a sufficently slow dependence on the spacetime coordinates $\bm{X}=(\bm{r},t)$. In fact, choosing $\bm{{\cal B}}(\bm{X})$ to give rise to spacetime textures allows to experimentaly carry out this new type of Berry-singularity tomography in a lock-in fashion~\cite{LockIn}, thus enabling the noise- and disorder-free detection of Berry singularities.

The engine of the here-proposed method relies on the construction of a quantity that I coin Berry sin\-gu\-la\-ri\-ty marker (BSM). The BSM enables the experimental diagnosis of the topological charges of the Berry singularities of the target Hamiltonian and their locations in the $d$-dimensional $\bm{K}=(\bm{k},M)$ space. Here, $\bm{k}$ denotes the wave vector, while $M$ is a parameter which tunes the size of the insulator's bulk energy gap and thus controls the occurence of topological phase transitions. The BSM is constructed as a suitable convolution of the re\-fe\-ren\-ce fields $(F_1,F_2,\ldots,F_d)$, and their expe\-ri\-men\-tal\-ly mea\-su\-ra\-ble conjugate fields $(P_1,P_2,\dots,P_d)$. In the micro\-ca\-no\-ni\-cal statistical ensemble, the conjugate fields are defined as $P_n=-\partial E/\partial F_n$, where $E$ denotes the ener\-gy of the sy\-stem and $n=1,2,3,\ldots,d$. As I detail in the remainder, the BSM tends by construction to a universal/topological value when $\bm{B}$ is such, so that the system approaches a topological phase transition. At the same time, the value of $\bm{B}$ required to observe the quantization of the BSM, also holds information regarding the location of the Berry sin\-gu\-la\-ri\-ties in $\bm{K}$ space. Thus, the BSM provides a measure for the distance away from the locations of the various Berry singularities and for this reason, it is naturally related to the so-called quantum or Fubini-Study metric~\cite{Provost} constructed using the Bloch band eigenvectors. Its value depends on the vector $\bm{B}$ and the modulus ${\cal B}(\bm{X})=|\bm{{\cal B}}(\bm{X})|$. The latter variable quantifies the re\-so\-lu\-tion of the experimental probe employed to perform the tomography proposed here. With no loss of ge\-ne\-ra\-li\-ty, I assume that ${\cal B}(\bm{X})$ is given by an $\bm{X}$-independent value which is from now on denoted ${\cal B}$.

The above construction guarantees that the BSM sa\-tisfies the following relation:
\bea
\lim_{{\cal B}\rightarrow0}{\rm BSM}(\bm{B}_s,{\cal B})=Q_s\,,
\eea

\noi where $Q_s$ defines the total topological charge of the Berry monopoles which me\-dia\-te the to\-po\-lo\-gi\-cal phase transition across $M=M_s$, and $\bm{B}_s$ is the value of $\bm{B}$ which is required for tuning the system at the $s$-th phase transition. For a sy\-stem lacking unitary symmetries, which is actually a prerequisite in the frame of the fundamental to\-po\-lo\-gi\-cal classification of insulators~\cite{Altland,SchnyderClassi,KitaevClassi,Ryu,TeoKane}, $Q_s$ stems from a single Berry singularity. In contrast, for a sy\-stem with additional unitary symmetries, the contributions of symmetry-related Berry singularities are not independent and, thus, one can in principle di\-sen\-tan\-gle the individual contribution of each one of the sin\-gu\-la\-ri\-ties.

In the following, I concretely exemplify how to construct the BSM for a representative AIII class topological insulator, which shares the same topological phe\-no\-me\-no\-lo\-gy with the Su-Schrieffer-Heeger (SSH) model~\cite{SSH}. Specifically, I examine the topological properties of this system in the presence of suitable external fields $\bm{F}$, demonstrate how the here-proposed expe\-ri\-men\-tal approach allows the detection of the under\-lying Berry singularities, and how the experimental measurements can be rendered quite insensitive to the presence of disorder.

The remainder is organized as follows. Section~\ref{sec:SectionII} describes the model Hamiltonian of the AIII topological insulator that I employ to demonstrate the here-proposed method. In Sec.~\ref{sec:SectionIII} I discuss the coupling of the electrons to the external background fields and the synthetic-space mapping of the Berry singularities that these achieve. Sections~\ref{sec:SectionIV} and~\ref{sec:SectionV} focus on the response of the insulator to these external fields. These sections respectively discuss the properties of the conjugate fields and the genera\-li\-zed su\-scepti\-bi\-li\-ties. Section~\ref{sec:SectionVI} contains the construction of the BSM, while Sec.\ref{sec:SectionVII} reveals the link between the BSM and the quantum metric tensor that characterizes the valence band. The next two sections, i.e., Secs.~\ref{sec:SectionVIII} and~\ref{sec:SectionIX} focus on the implementation of the BSM using synthetic textures. Section~\ref{sec:SectionX} contains an analysis of the effects of random disorder on the BSM, and shows how the BSM can be shielded against noise by employing the lock-in principle. Finally, Appendices~\ref{App:AppendixA}-\ref{App:AppendixB} discusse alternative possibi\-li\-ties regarding the coupling to the external fields and further useful technical details.

\section{Model Hamiltonian}\label{sec:SectionII} 

The approach proposed in this manuscript is demonstrated for the va\-riant of the SSH model which is defined in terms of the single-particle Bloch Hamiltonian $\hat{H}(\bm{K})=\bm{d}(\bm{K})\cdot\bm{\tau}$, with the ``isospin'' vector:
\begin{align}
\bm{d}(\bm{K})=\big(\sin\big(k-k_0\big),1-\cos k+M,0\big)\,.
\label{eq:gVector}
\end{align}

\noi The wavenumber $k$, as well as the offset $k_0$, are defined in the Brillouin zone (BZ) $[-\pi,\pi)$. To represent the above $2\times2$ matrix Hamiltonian I employ the ``isospin'' $\tau_{1,2,3}$ Pauli matrices and the respective unit matrix $\mathds{1}_{\tau}$. While the precise nature of the $\tau$ degree of freedom is irre\-le\-vant for the general idea behind the approach I present, this work is yet restricted to strictly non-supercon\-duc\-ting sy\-stems. The above Hamiltonian possesses a chiral symmetry effected by $\Pi=\tau_3$, which esta\-bli\-shes AIII as the ensuing symmetry class for the topological insulator. With no loss of generality, I set $k_0=0$ in the remainder of this work~\cite{CommentBDI}.

For the va\-lues $M\neq\{-2,0\}$, the system described by Eq.~\eqref{eq:gVector} features a full gap in the bulk energy spectrum and resides in the topologically-nontrivial regime for $M\in(-2,0)$, where it supports a single fractional zero-energy mode (FZM) at each edge of the system~\cite{JR}. The cha\-ra\-cte\-ri\-stic feature of these FZMs is that they possess a fixed chirality $\pm1$, i.e., the FZM eigenvectors are eigenstates of $\Pi$. Even more, FZMs on opposite edges see opposite chiralities. Hence, the fractionalization in the present case refers to the halving of the chirality degree of freedom experienced by the zero-energy modes. 

The topological properties and the absence/emergence of FZMs in the insulating phases occuring in the regime $M\in(\--\infty,-2)\cup(-2,0)\cup(0,\infty)$ are predicted by eva\-lua\-ting the winding number~\cite{SchnyderClassi}:
\begin{align}
w=\int_{\rm BZ}\frac{dk}{2\pi}\sum_{n,m}\varepsilon_{nm}\hat{d}_n(\bm{K})\frac{\partial\hat{d}_m(\bm{\bm{K}})}{\partial k}
\label{eq:WindingNumber}
\end{align}

\noi where $\varepsilon_{nm}$ denotes the antisymmetric Levi-Civita symbol with $n,m=1,2$, while $\hat{\bm{d}}(\bm{K})=\bm{d}(\bm{K})/|\bm{d}(\bm{K})|$ corresponds to the unit vector of $\bm{d}(\bm{K})$. As it has been shown in Ref.~\onlinecite{WeiChen1,WeiChen2}, the above topological invariant can also form the basis for constructing quantities which can cha\-rac\-te\-ri\-ze the topological phase transition using methods analogous to those used for studying critical phenomena.

Here, however, I am interested in a clearly different approach, which is motivated by the fact that the difference of $w$ across a bulk topological phase transition occu\-ring at $M=M_s$, which is defined as:
\begin{align}
\Delta w_s=w\big(M=M_s^+\big)-w\big(M=M_s^-\big)\label{eq:WindingDifference}
\end{align}

\begin{figure}[t!]
\begin{center}
\includegraphics[width=\columnwidth]{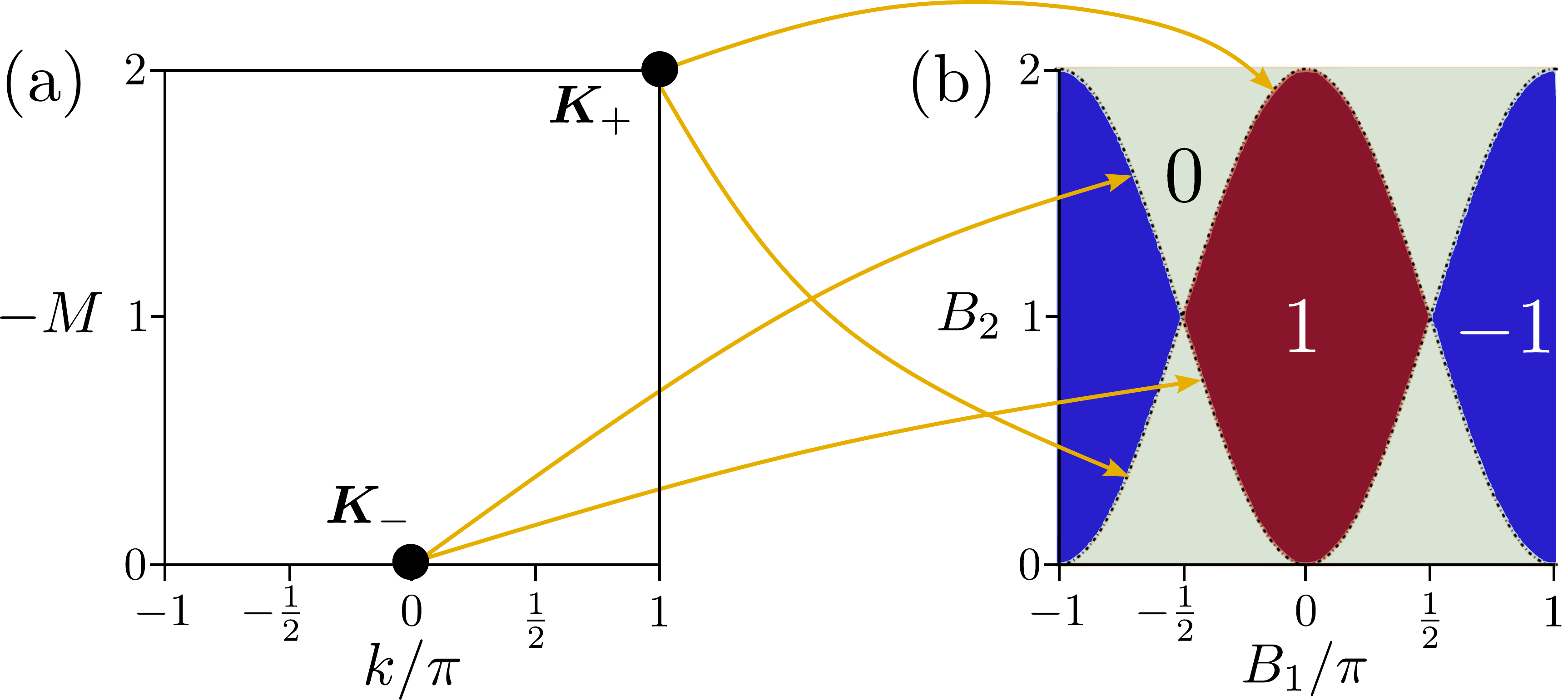}
\end{center}
\caption{Panel (a) depicts the two Berry singularities obtained from Eq.~\eqref{eq:gVector} in $\bm{K}=(k,M)$ space for $k_0=0$. These carry topological charge $\pm1$ and are located at $\bm{K}_+=(\pi,-2)$ and $\bm{K}_-=(0,0)$, respectively. (b) Topological phase diagram of model in Eq.~\eqref{eq:gVectorF} in the synthetic $\bm{B}$ space. The numbers shown correspond to the outcome of the winding number $w$ for $k_0=M=0$. The Berry singularities are mapped from $\bm{K}$ to $\bm{B}$ space as indicated by the arrows. Note that the mapping of the topological charges may differ up to an overall sign depending on the quadrant in synthetic space. The Berry vortices in $\bm{K}$ space also map to the remaining parts of the lines obtained by $B_1\mapsto-B_1$, which are not shown by arrows.}
\label{fig:Figure1}
\end{figure}

\noi yields the \textit{total} to\-po\-lo\-gi\-cal charge $Q_s$ of the Berry monopoles of $\hat{H}(\bm{K})$ which are responsible for the transition. Note that, in the present case, the Berry singularities correspond to vortices in 2D $\bm{K}=(k,M)$ space, which appear at the locations $\bm{K}_s=\big\{(\pi,-2),(0,0)\big\}$ with $s=\pm$. The two Berry vortices are depicted in Fig.~\ref{fig:Figure1}(a).

The charge of each Berry vortex can be equivalently obtained by means of the line integral:
\begin{align}
Q_s=-\ointctrclockwise_{{\cal C}_{\bm{K}_s}}\frac{d\bm{K}}{2\pi}\cdot\sum_{n,m}\varepsilon_{nm}\hat{d}_n(\bm{K})\frac{\partial\hat{d}_m(\bm{\bm{K}})}{\partial\bm{K}}\,,
\label{eq:Vorticity}
\end{align}

\noi where ${\cal C}_{\bm{K}_s}$ is a loop which encircles the singularity located at $\bm{K}_s$ and $n,m=1,2$. In fact, Eq.~\eqref{eq:WindingDifference} originates from Eq.~\eqref{eq:Vorticity} after a particular path ${\cal C}_{\bm{K}_s}$ has been chosen. 

The topological charge can be alternatively expressed in terms of the Berry vector potential $\bm{A}_\pm(\bm{K})$ of the Bloch eigenstates $\big|\bm{u}_\pm(\bm{K})\big>$ correspon\-ding to the energy dispersions $E_\pm(\bm{K})=\pm|\bm{d}(\bm{K})|$. In the present case, the Berry vector potential is the same for both bands and is defined as: 
\begin{align}
\bm{A}(\bm{K})=i\big<\bm{u}_-(\bm{K})\big|\partial_{\bm{K}}\bm{u}_-(\bm{K})\big>
=-\frac{1}{2}\varepsilon_{nm}\hat{d}_n(\bm{K})\frac{\partial\hat{d}_m(\bm{\bm{K}})}{\partial\bm{K}}
\end{align}

\noi which allows to rewrite $Q_s$ in the compact form:
\begin{align}
Q_s=\frac{1}{\pi}\ointctrclockwise_{{\cal C}_{\bm{K}_s}}d\bm{K}\cdot\bm{A}(\bm{K})\,.
\label{eq:VorticityA}
\end{align}

\noi From either one of the above equivalent definitions, one finds that the Berry vorti\-ci\-ties across the two transitions at $M_\pm=\{-2,0\}$ are correspondingly $Q_\pm=\{1,-1\}$.

Being in a position to know the topological charges and locations of the Berry singularities allows us to reconstruct the entire topological phase diagram of the sy\-stem, as long as the topological pro\-per\-ties of the sy\-stem are known for a single value of $M$. For convenience, this value can be associated with a well-known limiting topological behavior of the sy\-stem. For instance, for the vector $\bm{d}(\bm{K})$ considered in Eq.~\eqref{eq:gVector}, the limits $M\rightarrow\pm\infty$ lead to a topologically-trivial system, which is a piece of information that is typically accessible in experiments.

\section{Berry-singularity mapping to synthetic space}\label{sec:SectionIII}

In this paragraph, I proceed by describing the coupling of the sy\-stem to the external fields $\bm{F}$. Within the throughout-adopted slow spacetime variation of these fields, I assume that these couple to the system and ge\-ne\-ra\-te an equally slow spatiotemporal va\-ria\-tion for $\bm{K}$, such that near the Berry singularities we have $d\bm{K}/d\bm{X}\sim d\bm{F}/d\bm{X}=d\bm{{\cal B}}/d\bm{X}$. For the model of Eq.~\eqref{eq:gVector}, I first consider the following concrete type of coupling (see also the comment in Ref.~\onlinecite{CommentCoupling} and the discussion in App.~\ref{App:AppendixA}):
\begin{align}
\bm{d}(\bm{K},\bm{F})=\big(\sin\big(k-F_1\big),1-\cos k+M-F_2,0\big),
\label{eq:gVectorF}
\end{align}

\noi which gives rise to the accordingly modified energy di\-sper\-sions that I denote $E_\pm[\bm{K},\bm{F}(\bm{X})]=\pm|\bm{d}(\bm{K},\bm{F}(\bm{X}))|$.

Within the semiclassical picture adopted here, the presence of the fields allow us to map the Berry singu\-la\-ri\-ties from $\bm{K}$ space to $\bm{F}$ space. The Berry singularities appear in $\bm{K}$ space at the two $\bm{F}$-dependent locations:
\bea
\bm{K}_+(\bm{F})&=&\big(\pi+F_1,F_2-\cos F_1-1\big)\,,\no\\
\bm{K}_-(\bm{F})&=&\big(F_1,F_2+\cos F_1-1\big)\,.
\eea

\noi By momentarily considering ${\cal B}=0$, the above results imply that by solely restricting to $\bm{B}$ space, the two Berry singularities appear respectively along the trajectories:
\begin{align} 
B_2=1\pm\cos B_1+M\,.\label{eq:BStrajectories}
\end{align}

\noi These two Berry singularity branches are overlaid with dot-dashed lines on the topological phase diagram presented in Fig.~\ref{fig:Figure1}(b). The latter is defined solely in $\bm{B}$ space and is obtained under the assumption $M={\cal B}=0$.

The above results imply that, by introducing the external fields, we are in a position to map out the topological phase diagram of the system, even without knowing the precise form of the Hamiltonian governing it. While the importance of inferring the Hamiltonian is certainly indisputable, in actual experiments the priority is to be able to manipulate the properties of the system using the external parameters. Hence, from this point of view, mapping the Berry singularities onto the synthetic space spanned by the control fields is more important when it comes to potential applications. Evenmore, ``transferring'' the Berry singularities to a synthetic space has advantages, since there, the Berry singularities are present in an extended window and form lines, thus rendering this approach preferred for their study.

To this end, I stress that the role played by the synthetic space in this work and the purpose for its use, are clearly different than the ones of a number of recent stu\-dies regarding conventional~\cite{Riwar,Eriksson,Meyer_PRL,LevchenkoI,LevchenkoII,WeylCircuits} and topological~\cite{Balseiro,MeyerHouzet,PKsynthetic,MTMPRB,Sakurai} superconductors, which discuss topological phases and Berry singularities in a synthetic space. There, the topological properties are associated with a Hamiltonian which depends on the synthetic space coordinates. Here instead, the focus is on the parametric dependence of the conjugate fields, which constitute expectation values of the Hamiltonian of inte\-rest. The goal is to employ the external fields as a means to manipulate the response of the system and infer the topological properties of the underlying Hamiltonian.

\begin{figure*}[t!]
\begin{center}
\includegraphics[width=1.0\textwidth]{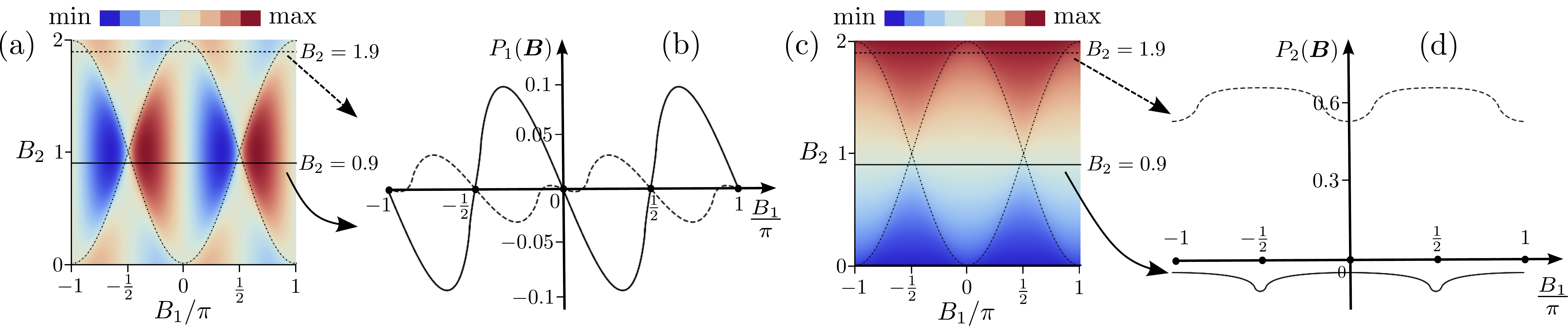}
\end{center}
\caption{Numerical evaluation of $\bm{P}(\bm{B})$, which is defined by setting ${\cal B}=0$ in $\bm{P}(\bm{F})$. Panels (a) and (c) depict a heat map of $P_{1,2}(\bm{B})$ as functions of $(B_1,B_2)$. The latter are sampled in the most interesting parameter-value window, in which Berry singularity branches exist. In (a) and (c), the colorcoding varies continuously between deep red to deep blue (see color bar), which one-to-one correspond to the maximum and minimum of the depicted function. In both panels, I have superimposed the contour plots $B_2\pm\cos B_1=1$ which yield the locations of the Berry singularities in the synthetic $\bm{B}$ space for $M=0$, which is the value I considered for the calculations presented here. Panels (b) and (d) show the dependence of $P_1(\bm{B})$ and $P_2(\bm{B})$, respectively, for two specific cuts of (a) and (c). These two cuts are for $B_2=1.9$ and $B_2=0.9$.}
\label{fig:Figure2}
\end{figure*}

\section{Response to the external fields}\label{sec:SectionIV}

The above method of imprinting the Berry singularities onto a synthetic space allows for their identification through the study of the response of the system to these external fields. As announced earlier, I consider external fields which consist of two components, the offset $\bm{B}$ and the spatio/temporal varying $\bm{{\cal B}}(\bm{X})$. The role of $\bm{B}$ is to control the topological properties of the system, and tune it near a topological phase transition, as it was demonstrated in the previous section. Being in a position to controllably approach a topological phase transition, at the same time allows us to tune the system near the location of a Berry sin\-gu\-la\-ri\-ty in synthetic space. On the other hand, the additional $\bm{{\cal B}}(\bm{X})$ component functions as a weak perturbation which is employed to assert the pre\-sen\-ce of the Berry sin\-gu\-la\-ri\-ty and, hence the topological character of the associated phase transition. The introduction of $\bm{{\cal B}}(\bm{X})$ also opens the door for a lock-in type of experimental detection of the Berry singularity charge. As I detail later on, this lock-in approach provides a route to filter out the effects of noise and disorder from the measurements.

The response of the system is encoded in the conjugate fields $\bm{P}[\bm{F}(\bm{X})]$. These are measured under fully/quasi-equilibrium conditions, since $\bm{F}(\bm{X})$ is here restricted to vary adiabatically in time and space. In the zero-temperature limit of interest, these receive contributions only from the occupied band of the Hamiltonian defined by Eq.~\eqref{eq:gVectorF}. The above conditions lead to the expression:
\bea
P_n[\bm{F}(\bm{X})]&=&-\frac{\partial}{\partial F_n(\bm{X})}\int_{\rm BZ}\frac{dk}{2\pi}\ph E_-[\bm{K},\bm{F}(\bm{X})]\no\\
&=&\int_{\rm BZ}\frac{dk}{2\pi}\ph\hat{\bm{d}}(\bm{K},\bm{F}(\bm{X}))\cdot\frac{\partial\bm{d}(\bm{K},\bm{F}(\bm{X}))}{\partial F_n(\bm{X})}.\qquad
\eea

To gain further insight regarding the properties of the system, I set ${\cal B}=0$ and calculate the conjugate fields as functions of $B_{1,2}$. For convenience, I assume $M=0$, and restrict to the regime $B_1\in[-\pi,\pi)$ and $B_2\in[0,2]$, where Berry singularities become accessible in the respective synthetic $\bm{B}$ space. From the panels of Fig.~\ref{fig:Figure2}, one observes that the vector $\bm{P}(\bm{B})$ and the synthetic coor\-di\-na\-tes $\bm{B}$, present an analogy to $\bm{d}(\bm{K})$ and $\bm{K}$, when it comes to their symmetry properties. Hence, based on the correspondence $\bm{d}(\bm{K})\mapsto\bm{P}(\bm{B})$, one expects that similar to $\bm{d}(\bm{K})$, which can be employed to define a nonzero vorticity in $\bm{K}$ space, the conjugate vector field $\bm{P}(\bm{B})$ can be employed to define vorticity in $\bm{B}$ space. However, such a construction is not possible since, in contrast to $|\bm{d}(\bm{K},\bm{B})|$, the modulus $|\bm{P}(\bm{B})|$ does not ne\-ce\-ssa\-ri\-ly contain zeros, and if it does, these do not have to coincide with the zeros of $|\bm{d}(\bm{K},\bm{B})|$ when restricted to $\bm{B}$ space. Nevertheless, the behaviour of $\bm{P}(\bm{B})$ is significantly influenced by the presence of the Berry singularities in synthetic space, since the variations of $\bm{P}(\bm{B})$ closely track the lines of the synthetic space Berry singularities given by Eq.~\eqref{eq:BStrajectories}. It is exactly this property which allows us to identify the presence of Berry sin\-gu\-la\-ri\-ties by extracting information about the variations of $\bm{P}(\bm{B})$.

\section{Perturbations about the Berry singularities}\label{sec:SectionV}

As mentioned above, one cannot directly use the conjugate fields $\bm{P}(\bm{B})$ or similarly $\bm{P}(\bm{F})$ to construct a topological invariant which can be associated with the Berry singularities in synthetic $\bm{F}$ space and, in turn, in $\bm{K}$ space. Nonetheless, the conjugate fields still hold information regarding the position and the charge of the Berry singularities in the synthetic space. This information can be accessed by de\-com\-po\-sing the field $\bm{F}$ into the two contributions $\bm{B}$ and $\bm{{\cal B}}$, with the latter containing the slow spatiotemporal variations mentioned earlier.

The crucial observation here is that when ${\cal B}\rightarrow0$, the response of the system can be retrieved by expan\-ding the conjugate fields up to first order in $\bm{{\cal B}}$. Such an expansion leads to the expression:
\begin{align}
P_n[\bm{F}(\bm{X})]\approx P_n(\bm{B})+\chi_{nm}(\bm{B}){\cal B}_m(\bm{X})
\end{align}

\noi where I introduced the generalized susceptibilities:
\begin{align}
\chi_{nm}(\bm{B})=\left.\frac{\partial P_n(\bm{F})}{\partial F_m}\right|_{{\cal B}=0},
\end{align}

\noi which in the present case obtain the following form: 
\bea
\chi_{nm}(\bm{B})&=&\int_{\rm BZ}\frac{dk}{2\pi}\ph\bigg[\frac{\partial\hat{\bm{d}}(\bm{K},\bm{B})}{\partial B_m}\cdot\frac{\partial\bm{d}(\bm{K},\bm{B})}{\partial B_n}\no\\
&&\qquad\qquad+\hat{\bm{d}}(\bm{K},\bm{B})\cdot\frac{\partial^2\bm{d}(\bm{K},\bm{B})}{\partial B_m\partial B_n}\bigg]\,.
\label{eq:SuscDef}
\eea

\noi From the above, one also obtains the equality relation $\chi_{21}(\bm{B})=\chi_{12}(\bm{B})$ for the off-diagonal tensor elements, since Eq.~\eqref{eq:SuscDef} is symmetric under the exchange $1\leftrightarrow2$.

\begin{figure}[t!]
\begin{center}
\includegraphics[width=1.0\columnwidth]{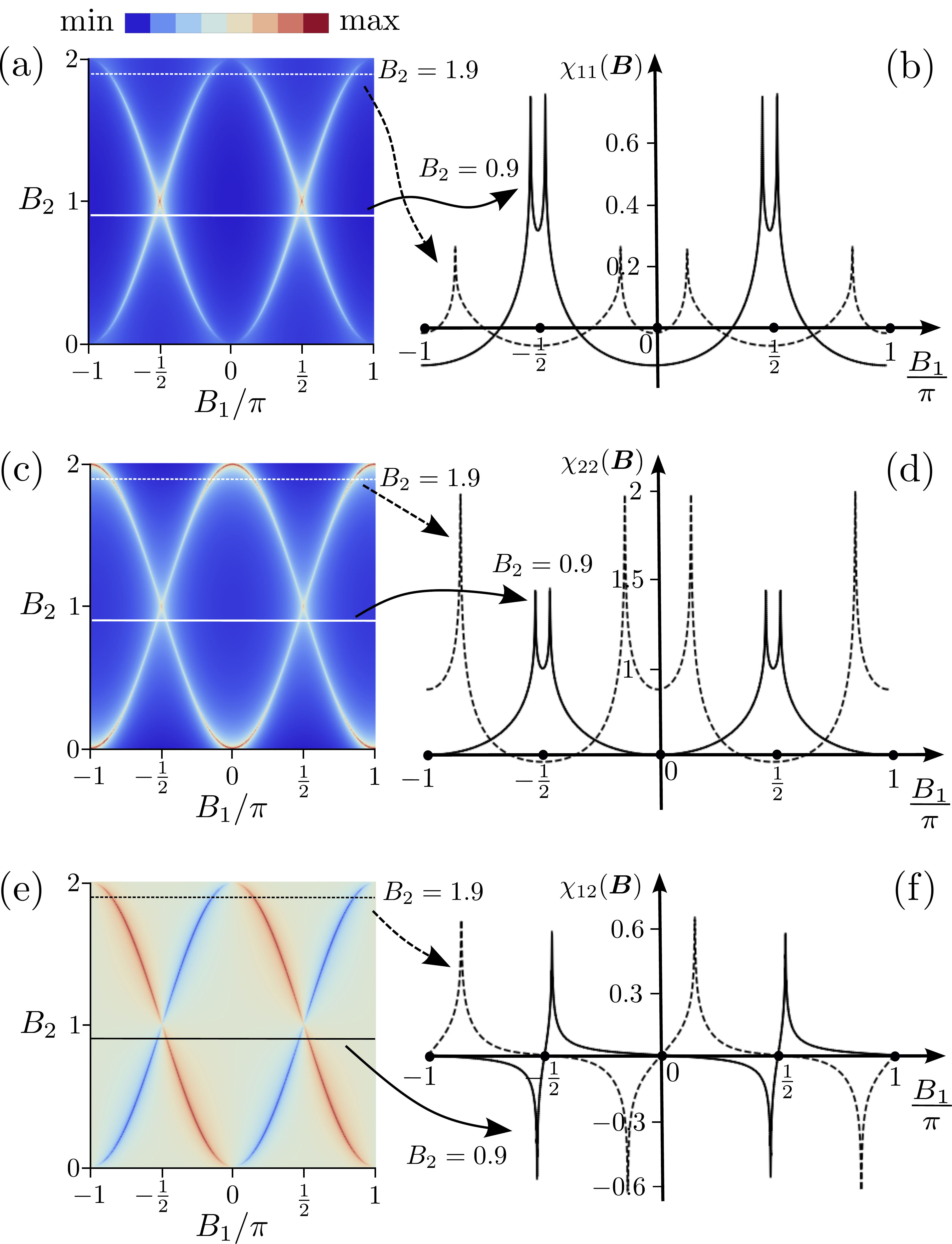}
\end{center}
\caption{Depiction of the numerically evaluated susceptibility tensor elements $\chi_{11,22,12}(\bm{B})$. Note that $\chi_{21}(\bm{B})=\chi_{12}(\bm{B})$. Panels (a), (c) and (e) present heat maps of the susceptibilities as functions of $(B_1,B_2)$. The paramerer window coin\-ci\-des with the one considered in Fig.~\ref{fig:Figure2}. Panels (b), (d) and (f) show the dependence of $\chi_{11,22,12}(\bm{B})$, for two specific cuts of (a), (c) and (e), respectively. These two cuts are also here for $B_2=1.9$ and $B_2=0.9$. In order to smoothen out the peaks appea\-ring at the singularities, I introduced to the Hamiltonian an additional term $d_3(\bm{K})\tau_3$ with a small nonzero component $d_3(\bm{K})=2\cdot10^{-3}$. This ``trick'' is employed in all calculations to follow, as long as these involve the evaluation of the susceptibility tensor elements. All the results are obtained for the value $M=0$.}
\label{fig:Figure3}
\end{figure}

In Fig.~\ref{fig:Figure3}, I present the outcome for each one of the three su\-sceptibility tensor ele\-ments $\chi_{11,22,12}(\bm{B})$. As assu\-med in the calculation concerning the conjugate fields, the value $M=0$ is considered also here. One observes that the three su\-sce\-pti\-bi\-li\-ties $\chi_{11,22,12}(\bm{B})$ become consi\-de\-ra\-ble mainly at the locations $\bm{B}_\pm$ of the Berry singulatities, which are given by $(B_1,B_2)=(B_1,1\pm\cos B_1)$. Hence, either su\-scep\-ti\-bi\-li\-ty can be employed to identify the two Berry singularity branches. In addition, we observe that $\chi_{11,22}(\bm{B})$ are positive in the vicinity of the Berry singularities. Instead, $\chi_{12}(\bm{B})$ presents sign changes which reflect the symmetry pro\-per\-ties of the sy\-stem, and follows a distinctive pattern when it comes to the two Berry singularity branches. In particular, we find that the two branches are cha\-ra\-cte\-ri\-zed by op\-po\-si\-te $\chi_{12}(\bm{B})$ which can be associated with the sign of the topological charge of the Berry singularities at $\bm{B}_+=(0,2)$ and $\bm{B}_-=(0,0)$, which map one-to-one the two $\bm{K}_\pm$ singularities in $\bm{K}$ space, as shown in Fig.~\ref{fig:Figure1}.

The connection between the structure of $\chi_{12}(\bm{B})$ in the $(B_1,B_2)$ plane and the topological charges of the Berry singularities in $\bm{K}$ space can be established by noticing that near the Berry singularities $(|B_1|/\pi,B_2)=(\nicefrac{1}{2},1)$, $\bm{d}(\bm{K},\bm{B})$ becomes linear in $\bm{K}-\bm{B}$, and can be expanded as $\bm{d}(\bm{K},\bm{B})\approx\big(\pm(k-B_1),1+M-B_2\big)$. This implies that the off-diagonal susceptibility now becomes:
\bea
\chi_{12}(\bm{B})\approx\int\frac{dk}{2\pi}\frac{d_1(\bm{K},\bm{B})d_2(\bm{K},\bm{B})}{|\bm{d}(\bm{K},\bm{B})|^3}\,.
\label{eq:Susc12Approx}
\eea

\noi The product $d_1(\bm{K},\bm{B})d_2(\bm{K},\bm{B})$ provides information about the order of the functions $d_{1,2}(\bm{K})$ in terms of $\bm{K}$, and thus allows us to indirectly infer the winding number of the vector $\hat{\bm{d}}(\bm{K})$, as well as the topological charges of the Berry sin\-gu\-la\-ri\-ties. From Fig.~\ref{fig:Figure3}, we find that in the vicinity of $\bm{B}=(B_1/\pi,B_2)=(\mp\nicefrac{1}{2},1)$ the off-diagonal susceptibility presents a quadrupolar spatial pattern $\sim \delta B_1\delta B_2$ in terms of the deviations $(\delta B_1,\delta B_2)$ of $(B_1,B_2)$ about these two Berry singularities which allow for the linearization of $\bm{d}(\bm{K},\bm{B})$. Such a quadrupolar pattern is characteristic of class AIII, in which Berry singula\-ri\-ties are nondegenerate and carry charge $\pm1$. Therefore, here, the sign of these quadrupolar profiles or simply the sign of $\chi_{12}(\bm{B})$, are sufficient to infer the topological charges~\cite{CommentHigher}. Notably, a similar method to determine the winding of a vector was earlier discussed by Kourtis in Ref.~\onlinecite{Kourtis}, where he proposed to identify the topological charge of Weyl points using polarization-resolved resonant inelastic x-ray scattering.

Although the off-diagonal su\-scepti\-bility provides all the crucial information that reveals the presence of a Berry singularity, i.e., (i) it peaks at the singularities and (ii) its sign is opposite for the two Berry singu\-la\-ri\-ty branches, it does not however appear as a prominent candidate to base the lock-in technique which I wish to propose in this work. The first reason is that $\chi_{12}(\bm{B})$ is practically zero or takes very small values in the largest part of the paramater regime, which can pose obstacles to its experimental detection. Motivated by this shortcoming, in the next section I provide a definition for a BSM which makes use of a non-negative quantity to describe the distance from a Berry sin\-gu\-la\-ri\-ty and, thus appears in principle more robust against disorder than $\chi_{12}(\bm{B})$.

\section{BSM Construction}\label{sec:SectionVI}

The above conclusions regarding the response of the system are the ones that strongly motivate the introduction of the additional field $\bm{{\cal B}}$. This is because upon their addition we are given the possibility to probe the elements of the susceptibility tensor as functions of $\bm{B}$. To make this transparent I study the properties of the fol\-lowing line integral:
\begin{align}
I_s(\bm{{\cal B}})=-\sum_{n,m}\ointctrclockwise_{{\cal C}_{\bm{B}_s}}d{\cal B}_n\varepsilon_{nm}P_m(\bm{B}_s+\bm{{\cal B}})\,.
\label{eq:Integral4BSM}
\end{align}

\noi The above construction is inspired by Eq.~\eqref{eq:VorticityA} and bears analogies to it. There, the line integral involves the Berry vector potential $\bm{A}(\bm{K})$ for the occupied band. Here, one aims at defining a quantity which is analogous to $\bm{A}(\bm{K})$. 

To achieve this goal, it is pivotal to investigate the behaviour of $I_s$ for a sufficiently small ${\cal B}=|\bm{{\cal B}}(\bm{X})|$, so that the system lies in the vicinity of a Berry singularity. In this event, $I_s$ is appro\-xi\-ma\-tely given by:
\begin{align}
I_s(\bm{{\cal B}})\approx-\sum_{n,m,\ell}\varepsilon_{nm}\chi_{m\ell}(\bm{B}_s)\ointctrclockwise_{{\cal C}_{\bm{B}_s}}d{\cal B}_n{\cal B}_\ell,
\label{eq:Integral4BSMapprox}
\end{align}

\noi since the contribution propotional to $P_n(\bm{B}_s)$ drops out after integrating over the closed path ${\cal C}_{\bm{B}_s}$. It is straightforward to evaluate the line integral. By considering for instance a circular path ${\cal C}_{\bm{B}_s}$, which is defined through setting $\bm{{\cal B}}={\cal B}(\cos\vartheta,\sin\vartheta)$ with $\vartheta\in[0,2\pi)$, one obtains:
\begin{align}
\ointctrclockwise_{{\cal C}_{\bm{B}_s}}d{\cal B}_n{\cal B}_\ell=\int_0^{2\pi}d\vartheta\ph\frac{d{\cal B}_n}{d\vartheta}\ph{\cal B}_\ell=-\pi{\cal B}^2\varepsilon_{n\ell}\,.
\label{eq:BIntegral}
\end{align}

\noi The above result implies that in the limit ${\cal B}\rightarrow0$, $I_s$ tends to the value:
\begin{align}
I_s({\cal B}\rightarrow0)\approx\pi{\cal B}^2\big[\chi_{11}(\bm{B}_s)+\chi_{22}(\bm{B}_s)\big]\equiv\pi{\cal B}^2{\rm Tr}\big[\check{\chi}(\bm{B}_s)\big],
\label{eq:Integral4BSMapproxLimit}
\end{align}

\noi where I introduced the generalized susceptibility tensor denoted here $\check{\chi}$, along with the trace operation in the space spanned by its indices. 

At this point I remind the reader that all the susceptibilities diverge at the Berry singularities branches, and in fact only at these. Therefore, since $I_s(\bm{{\cal B}})$ becomes substantial practically only near the $s$-th Berry sin\-gu\-la\-ri\-ty, it can supply us with a notion of a distance from this sin\-gu\-la\-ri\-ty. Indeed, as I discuss in the next section, this conclusion is further backed by the fact that $I_s$ relates to the quantum metric tensor $\check{g}$ which quantifies the di\-stance between two quantum states. Note that the utility of $\check{g}$ in asserting the occurence of phase transition has already been emphasized in previous works~\cite{Zanardi,HQLin,Ma1}. From a practical point of view, here, one can simply ``nor\-ma\-li\-ze'' $I_s(\bm{{\cal B}})$ for a vector $\bm{{\cal B}}$ of an arbitrary strength ${\cal B}$, by dividing it by $I_s({\cal B}\rightarrow0)$, in order to obtain a normalized distance from the Berry singularity and infer the location of the latter in synthetic space. 

Despite the fact that $I_s$ provides a measure of distance from Berry singularities, it nevertheless cannot provide information regarding the sign of the charge of the Berry singularities, since the sign of $I_s$ is fixed. To tackle this problem, I introduce a quantity which plays an analogous role to the Berry vector potential and is defined as:
\begin{align}
{\cal A}_n(\bm{B},\bm{{\cal B}})=-\frac{{\rm sgn}\left[\chi_{12}(\bm{B})\right]}{{\rm Tr}\big(\check{\chi}_{\rm BS}\big){\cal B}}\sum_m\varepsilon_{nm}P_m(\bm{B}+\bm{{\cal B}}),
\label{eq:pFields}
\end{align}

\noi where the sign of the susceptibility $\chi_{12}(\bm{B})$ is introduced in order to provide the so-far missing piece of information regarding the sign of the charge of the Berry singularity. 

The quantity ${\rm Tr}\big(\check{\chi}_{\rm BS}\big)$ denotes the value of ${\rm Tr}\big(\check{\chi}\big)$ eva\-lua\-ted at the location of the Berry singularity which is the closest to the synthetic space location $\bm{B}$ of interest. Such an approach is employed for the numerical studies in Sec.~\ref{sec:SectionIX}, and enables us to restrict the BSM to va\-lues in the window $[-1,1]$ for all Berry singularities of the branches $B_2=1\pm\cos B_1+M$.  Notably, in case the trace of the suscepti\-bi\-li\-ty is fully determined by the quantum metric tensor $\check{g}$, ${\rm Tr}\big(\check{\chi}_{\rm BS}\big)$ reaches a universal value in almost the entire region of synthetic space where Berry singularities emerge~\cite{CommentUniversal}. This holds when the diamagnetic contributions to the susceptibilities are either negligible or precisely zero~\cite{CommentSection}, as I clarify later.

Under the above conditions, and along the lines of the line integral constructed in Eq.~\eqref{eq:Integral4BSMapprox}, I define the BSM as follows:
\begin{align}
{\cal Q}(\bm{B},\bm{{\cal B}})=\frac{1}{\pi}\ointctrclockwise_{{\cal C}_{\bm{B}}}d\hat{\bm{{\cal B}}}\cdot\bm{{\cal A}}(\bm{B},\bm{{\cal B}})\,,
\label{eq:BSM}
\end{align}

\noi where $\hat{\bm{{\cal B}}}=\bm{{\cal B}}/{\cal B}$ and ${\cal C}_{\bm{B}}$ defines a closed path centered at location $\bm{B}$. Notably, the BSM tends to a quantized value for ${\cal B}\rightarrow0$, that is:
\begin{align}
\lim_{{\cal B}\rightarrow0}{\cal Q}_s(\bm{{\cal B}})={\rm sgn}[\chi_{12}(\bm{B}_s)]\,.
\label{eq:LimitBSM}
\end{align}

It is important to clarify that the construction of the BSM bears, up to some extent, similarities to the to\-po\-lo\-gi\-cal marker introduced by Bianco and Resta in Ref.~\onlinecite{Resta}, and further developed in Refs.~\onlinecite{MarkerCurrents},~\onlinecite{MarkerInteractions}, and others. The main common feature is that, under a suitable limit, the BSM becomes a topological invariant quantity similar to the topological marker. However, in stark contrast to the latter, the BSM is aimed at being constructed using experimentally-inferred quantities, while the topological marker is more suitable for predicting the local to\-po\-lo\-gi\-cal properties of the system from a given Hamiltonian.

Another important difference is that the BSM provides the charge of the Berry singularities across to\-po\-lo\-gical phase transitions, while the topological marker provides the local value of the topological invariant of the system. Finally, an additional distinctive feature is that the BSM constitutes a quantity which probes the spacetime nonlocality of the sy\-stem, since it is evaluated using a particular combination of va\-lues of $\bm{{\cal A}}[\bm{{\cal B}}(\bm{X})]$ obtained at different spacetime coordinates $\bm{X}$.\\

\section{Berry singularity marker and quantum geometry}
\label{sec:SectionVII}

The construction of the BSM relies on the presence of the quantity ${\rm Tr}\big[\check{\chi}(\bm{B})\big]$ whose properties remain to be investigated. As I argued earlier, this trace appears capable of quantifying the distance of the sy\-stem from a Berry singularity. Indeed, reaching to such a conclusion is justifiable since, at it emerges from previous works~\cite{Provost,Neupert,Zanardi} and I further clarify here, the susceptibility tensor $\check{\chi}$ relates to the quantum metric tensor $\check{g}$. 

\begin{figure*}[t!]
\begin{center}
\includegraphics[width=1\textwidth]{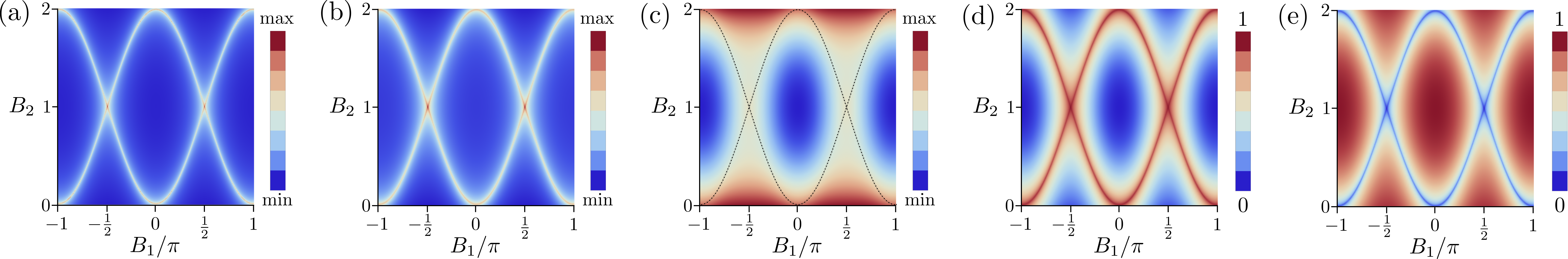}
\end{center}
\caption{Numerical results for (a) ${\rm Tr}\big(\check{\chi}\big)$, (b) ${\rm Tr}\big(\check{\chi}^{(\rm p)}\big)$, (c) ${\rm Tr}\big(\check{\chi}^{(\rm d)}\big)$, and (d/e) $|{\rm Tr}(\check{\chi}^{(\rm p/d)})|/\sqrt{[{\rm Tr}(\check{\chi}^{(\rm p)})]^2+[{\rm Tr}(\check{\chi}^{(\rm d)})]^2}$. From panels (a)-(b) one confirms that the paramagnetic/geometric contribution dominates the total susceptibility in the vi\-ci\-ni\-ty of the Berry singularity branches. Instead, the diamagnetic component shown in (c) is practically constant along these branches, which are overlaid with dot-dashed lines. Panels (d) and (e) present a comparison between the normalized paramagnetic and diamagnetic contributions, which confirm once again the dominance of the former near Berry singularities. In fact, the pa\-ra\-ma\-gne\-tic/geometric part is constant in almost the entire synthetic space. Deviation from this universal behavior is obtained near the points $\bm{B}=(\pm\pi/2,1)$ and $\bm{B}=(\pi,2)$ in which case the Hamiltonian cannot be linearized with respect to neither $\bm{K}$ nor $\bm{B}$. All the calculations are carried out for $M=0$.}
\label{fig:Figure4}
\end{figure*}

To demonstrate their precise connection, I re-express the susceptibility tensor through the eigenstates of the Hamiltonian, since these are employed for the construction of the quantum metric. Therefore, in terms of the adiabatic Bloch eigenstates $\big|\bm{u}_a(\bm{K},\bm{B})\big>$ correspon\-ding to the energy bands $E_a(\bm{K},\bm{B})$, $\check{\chi}$ is decomposed into the so-called paramagnetic $\check{\chi}^{({\rm p})}$ and diamagnetic parts $\check{\chi}^{({\rm d})}$. The paramagnetic contribution is a result of linear response of the system with vertices defined by the ge\-ne\-ra\-li\-zed pa\-ra\-ma\-gne\-tic currents for ${\cal B}=0$. In an operator form, these currents are given by the matrix elements:
\begin{align}
J_n^{ab}=-\big<\bm{u}_a\big|\partial_{B_n}\hat{H}\big|\bm{u}_b\big>
\equiv\big(E_a-E_b\big)\big<\bm{u}_a\big|\partial_{B_n}\bm{u}_b\big>,
\label{eq:GeneralizedCurrent}
\end{align}

\noi where the argument $(\bm{K},\bm{B})$ is implied in all the above functions. The above expression only accounts for interband elements with $a\neq b$, since we are here interested in the zero temperature response of an insulator. On the other hand, the diamagnetic contribution stems from the expectation value of the derivative of the above currents.

Straightforward calculations, reveal that the diagonal ele\-ments of the paramagnetic part read as: 
\begin{align}
\chi_{nn}^{(\rm p)}=\int_{\rm BZ}\frac{dk}{2\pi}\ph\sum_a
2\big<\partial_{B_n}\bm{u}_a\big|\big(\hat{H}-E_a\big)\big|\partial_{B_n}\bm{u}_a\big>n_F\big(E_a\big),\label{eq:ParamSusc}
\end{align}

\noi and their diamagnetic counterparts are obtained as:
\begin{align}
\chi_{nn}^{(\rm d)}=-\int_{\rm BZ}\frac{dk}{2\pi}\ph\sum_a\big<\bm{u}_a\big|\partial^2\hat{H}/\partial B_n^2\big|\bm{u}_a\big>n_F\big(E_a\big)\,,
\end{align}

\noi where I suppressed the various arguments for compact\-ness. In the above, $n_F(\epsilon)$ defines the Fermi-Dirac distri\-bution for an energy $\epsilon$, while the indices $a,b$ label the Bloch eigenvectors.  In the case of the model of Eq.~\eqref{eq:gVector}, the indices $a,b$ run over the values $\pm$, and the paramagnetic (diamagnetic) susceptibility presented above leads to the contribution of the first (second) row in  Eq.~\ref{eq:SuscDef}.

At zero temperature only the occupied band of energy $E_-(\bm{K},\bm{B})=-|\bm{d}(\bm{K},\bm{B})|$ contributes to the susce\-pti\-bi\-li\-ty. Hence, the expressions for the susceptibilities simplify further. By employing the projectors $\hat{{\cal P}}_a$ of the Bloch states $\big|\bm{u}_a\big>$, one arrives at the following expression for the paramagnetic contribution:
\begin{align}
\chi_{nn}^{(\rm p)}(\bm{B})=\int_{\rm BZ}\frac{dk}{2\pi}\ph
4|\bm{d}(\bm{K},\bm{B})|g_{nn}(\bm{K},\bm{B}),\label{eq:SuscGeom}
\end{align}

\noi where I introduced the diagonal components of the quantum metric tensor of the occupied band~\cite{CommentStiffness,Torma,LongLiang}
\begin{align}
g_{nn}=\big<\partial_{B_n}\bm{u}_-\big|\big(\hat{\mathds{1}}-\hat{{\cal P}}_-\big)\big|\partial_{B_n}\bm{u}_-\big>,
\end{align}

\noi which, in the present situation, is defined in terms of the synthetic coordinates $\bm{B}$, instead of the base space coordinates $\bm{K}$. The above definition of the diagonal quantum metric implies that the quantum distance $ds$ is given by the expression $ds^2=g_{nm}dB_ndB_m$ with $n,m=1,2$, where also the similarly-defined off-diagonal components of $\check{\chi}$ are involved. 

The demonstrated connection between $\check{g}$ and $\check{\chi}$ sheds light to the results of the previous section and in particular the limit obtained in Eq.~\eqref{eq:Integral4BSMapproxLimit}. Using the above definitions, the paramagnetic contribution of the latter equation is re-exressed in the following form:
\begin{align}
\frac{I_s^{(\rm p)}({\cal B}\rightarrow0)}{\pi}\approx{\cal B}^2\int_{\rm BZ}\frac{dk}{2\pi}\ph4|\bm{d}(\bm{K},\bm{B}_s)|{\rm Tr}\big[\check{g}(\bm{K},\bm{B}_s)\big]\,,\label{eq:Integral4BSMapproxLimitmetric}
\end{align}

\noi which is analogous to the result obtained in Ref.~\onlinecite{Neupert} to describe the relation between the power spectrum of fluctuations and the quantum geometry. The possi\-bi\-li\-ty of relating response functions to the quantum metric and geometric tensor in ge\-ne\-ral, as well as the connection to the fluctuations dissipation theorem had been already pointed out by Provost and Vallee in their seminal paper~\cite{Provost}. Here, I instead explore the zero-frequency response of the system, since I focus on band touchings.

Remarkably, near a Berry singularity, the emergent ``mi\-ni\-mal'' coupling to the external fields, that reads $\bm{K}\mapsto\bm{K}-\bm{B}$ or $\bm{K}\mapsto\bm{K}+\bm{B}$ depending on the location of the Berry singularity, implies that each dia\-go\-nal element of the quantum metric $g_{nn}$ becomes approximately equal to the squared modulus of the respective component $\big[A_n(\bm{K},\bm{B}_s)\big]^2$ of the Berry vector potential, where I also inserted the dependence of the Berry vector potential on the added external field $\bm{B}$. This becomes apparent from the expression of the quantum metric tensor for the occupied band which reads as:
\bea
g_{nn}(\bm{K},\bm{B})&=&\left[\frac{1}{2}\ph\hat{\bm{d}}(\bm{K},\bm{B})\times\partial_{B_n}\hat{\bm{d}}(\bm{K},\bm{B})\right]^2\\
&\equiv&\frac{\big[\bm{d}(\bm{K},\bm{B})\times\partial_{B_n}\bm{d}(\bm{K},\bm{B})\big]^2}{4|\bm{d}(\bm{K},\bm{B})|^4}\,.
\eea

\noi From the above, one concludes that the trace of the quantum metric encodes information regarding the Berry vor\-ti\-ci\-ty density, i.e., the integrand of Eq.~\ref{eq:Vorticity}, which blows up when the system is tuned at the Berry sin\-gu\-la\-ri\-ty. 

Even more, since near each Berry sin\-gu\-la\-ri\-ty the Hamiltonian can be linearized with respect to $\bm{K}$, and concomitantly $\bm{B}$, the diamagnetic contribution of each Berry singularity to the susceptibility is negligible. Therefore, va\-ria\-tions of the susceptibility in the vi\-ci\-ni\-ty of the Berry singularity branches are mainly dictated by its geometric contribution, which gets simplified in the following fashion:
\begin{align} 
{\rm Tr}\big[\check{\chi}^{(\rm p)}(\bm{B}\rightarrow\bm{B}_s)\big]\rightarrow
\int_{\rm BZ}\frac{dk}{2\pi}\frac{1}{|\bm{d}(\bm{K},\bm{B}_s)|}\,.\end{align}

\noi In fact, the above geometric contribution ${\rm Tr}\big[\check{\chi}(\bm{B}_s)\big]$ is expected to take the same value at all the positions in $\bm{B}$ space which correspond to Berry sin\-gu\-la\-ri\-ties. Note, however, that this holds as long as the Hamiltonian is linea\-ri\-za\-ble in terms of both fields $B_{1,2}$. If this requirement is not met for some value of $\bm{B}_s$, the geometric contribution will deviate from this universal value.

The abovementioned properties of the elements of the susceptibility tensor are confirmable by means of a quantitative exploration of the trace of the total, pa\-ra\-ma\-gne\-tic, and diamagnetic susceptiblities in the case of the model interest. Pa\-nels (a)-(c) of Fig.~\ref{fig:Figure4} depict the three types of susceptibility traces in the sequence mentioned. One observes that the total susceptibility is mainly determined by its geo\-me\-tric part in the vicinity of the Berry sin\-gu\-la\-ri\-ty branches, where the diamagnetic part remains practically constant. Panels (d) and (e) present the normalized paramagnetic and diamagnetic contributions. Indeed the geometric part dominates near the Berry singularities. Even more, its heat map features a higher contrast, thus implying that it shows more abrupt variations compared to its diamagnetic counterpart.

Another important aspect is that the geometric component of the trace takes almost the same value for almost the entire part of the Berry singularity branches, with the only exceptions being the neighborhoods of $\bm{B}=(\pm\pi/2,1)$ and $\bm{B}=(\pi,2)$. The reason is that for these values, the term $\sin k$ or $\cos k$ appearing in $\bm{d}(\bm{K},\bm{B})$ does not contain a linear term $\propto k$ in its Taylor expansion about the gap-closing wavenumbers $k=\{\pm\pi/2,\pi\}$.

Noteworthy, the above universal value of the geo\-me\-tric component of the trace at the Berry singularities, also coincides with the respective value of the chiral su\-scep\-tibility $\chi_{\Pi}(\bm{B})$. The latter is defined as the suscepti\-bi\-li\-ty obtained in response to the perturbation entering the Hamiltonian via the term $-m\Pi$, where $m$ plays the role of an external field breaking chiral symmetry. Cal\-cu\-la\-ting this susceptibility in the limit $m\rightarrow0$, leads to the analytical expression:
\begin{align}
\chi_{\Pi}(\bm{B})=2\int_{\rm BZ}\frac{dk}{2\pi}\ph\sum_{a,b}
\big|\big<\bm{u}_a\big|\Pi\big|\bm{u}_b\big>\big|^2\frac{n_F\big(E_a\big)}{E_b-E_a}\,.
\end{align}

\noi Using $\Pi\big|\bm{u}_a\big>=\big|\bm{u}_{-a}\big>$ and $E_{-a}=-E_a$, one finds that in the present case the above expression takes the form:
\begin{align}
\chi_{\Pi}(\bm{B})
=\int_{\rm BZ}\frac{dk}{2\pi}
\frac{1}{|\bm{d}(\bm{K},\bm{B})|}\,.
\end{align}

The above result further implies that, when the diamagnetic response is negligible, one can use $\chi_{\Pi}(\bm{B})$ as the normalization factor ${\rm Tr}\big(\check{\chi}_{\rm BS}\big)$ in Eq.~\eqref{eq:pFields}. While employing one or the other normalization method does not lead to substantial differences in regards to the theo\-re\-ti\-cal analysis, using $\chi_{\Pi}(\bm{B})$ as the normalization factor may be preferrable in certain experimental situations, in which, the measurement of the expectation value of the chiral symmetry operator $\Pi$ is easily accessible. In fact, one can pursue calculating the BSM using both nor\-ma\-li\-za\-tion approaches, since this can allow to corroborate the chiral-symmetry properties of the system, and estimate the contribution of the diamagnetic currents.

\section{Implementation of the BSM using Spacetime Textures}\label{sec:SectionVIII}

The definition of the BSM in Eq.~\eqref{eq:BSM} requires the implementation of a closed loop in synthetic space, which can be generated by considering that $\bm{{\cal B}}$ varies in space and/or time in a pe\-rio\-dic fashion. How to choose the dependence of $\bm{{\cal B}}$ on $\bm{X}$ is decided based on the expected topological pro\-per\-ties of the target sy\-stem. Specifically, $\bm{{\cal B}}(\bm{X})$ is chosen to give rise to a topological texture cha\-ra\-cte\-ri\-zed by the same kind of topological charge which is expected to classify the Berry singularities of the target bulk Hamiltonian. If little is known about the latter, then one has to proceed by testing various spacetime profiles in order to exhaust all the topological classes that are possible in the given spacetime dimensionality~\cite{Altland,SchnyderClassi,KitaevClassi,Ryu,TeoKane}.

For the situation of interest, the desired stra\-te\-gy is to consider spatial/spacetime texture profiles for $\bm{{\cal B}}(\bm{X})$, which can yield a nonzero spacetime vorticity defined as:
\begin{align}
Q_{\bm{{\cal B}}}=-\frac{1}{2\pi}\ointctrclockwise_{{\cal C}_{\bm{X}}}d\bm{X}\cdot\sum_{n,m}\varepsilon_{nm}\hat{{\cal B}}_n(\bm{X})\frac{\partial\hat{{\cal B}}_m(\bm{\bm{X}})}{\partial\bm{X}}
\label{eq:TextureVorticity}
\end{align}

\begin{figure}[t!]
\begin{center}
\includegraphics[width=\columnwidth]{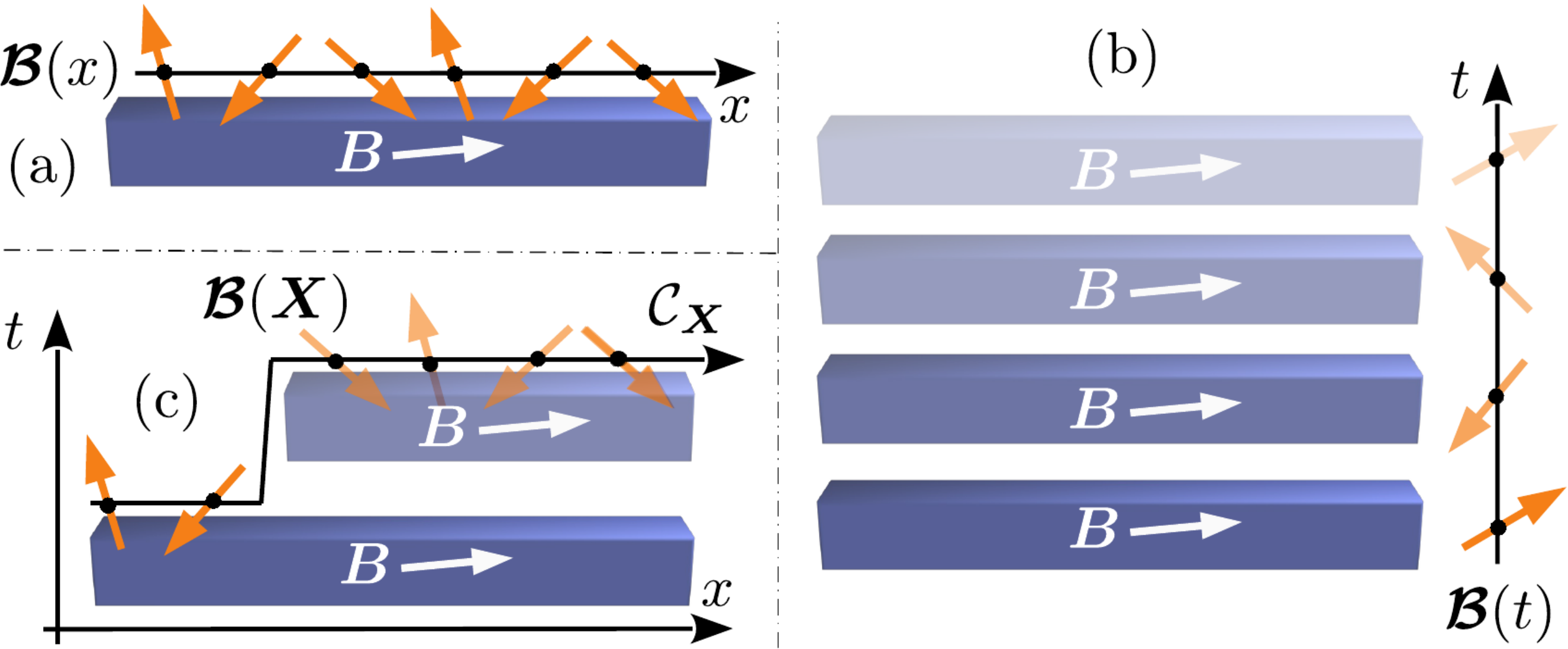}
\end{center}
\caption{Three different ways to generate a spacetime texture for the perturbation $\bm{{\cal B}}(\bm{X})$. One can employ a spiral defined solely in spatial/time coordinates, as shown in (a)/(b), or a spiral unfolding along a spacetime line ${\cal C}_{\bm{X}}$. To ensure that ${\cal C}_{\bm{X}}$ is equivalent to a circle $\mathbb{S}^1$, the values for $\bm{{\cal B}}(\bm{X})$ are required to be identical at the beginning and end of the path.}
\label{fig:Figure5}
\end{figure}

\noi with ${\cal C}_{\bm{X}}$ a spacetime loop encircling the vortex contained in the $\bm{{\cal B}}(\bm{X})$ field. Also here, the indices read $n,m=1,2$ since $\bm{X}=(X_1,X_2)=(x,t)$ spans a Euclidean metric space within the semiclassical picture adopted throughout this work. See Fig.~\ref{fig:Figure5} for a schematic of possible types of such spacetime vortices.

By exploiting the general connection between the winding number and the vorticity, it is convenient to assume a texture of the form $\bm{{\cal B}}(\vartheta)={\cal B}\big(\cos(\nu\vartheta),\sin(\nu\vartheta)\big)$ with $\vartheta\in[0,2\pi)$, which describes simultaneously a circle $\mathbb{S}^1$ or a compactifiable line along any direction in 2D spacetime $\bm{X}$. The above texture gives rise to a nonzero winding number $w_{\bm{{\cal B}}}=\nu\in\mathbb{Z}$, which is defined through the expression:
\begin{align}
w_{\bm{{\cal B}}}=\int_0^{2\pi}\frac{d\vartheta}{2\pi}\sum_{n,m}\varepsilon_{nm}\hat{{\cal B}}_n(\vartheta)\frac{d\hat{{\cal B}}_m(\vartheta)}{d\vartheta}\,.
\label{eq:TextureWinding}
\end{align}

With the help of the above, I now write the BSM defined in Eq.~\eqref{eq:BSM} as follows:
\bea
{\cal Q}(\bm{B},{\cal B})&=&\frac{1}{\pi}\int_0^{2\pi}d\vartheta\ph\frac{d\hat{\bm{{\cal B}}}(\vartheta)}{d\vartheta}\cdot\bm{{\cal A}}[\bm{B},{\cal B}(\vartheta)]\no\\
&\equiv&
-\frac{1}{\pi}\int_0^{2\pi}d\vartheta\ph\hat{\bm{{\cal B}}}(\vartheta)\cdot\frac{d\bm{{\cal A}}[\bm{B},{\cal B}(\vartheta)]}{d\vartheta}\,.
\label{eq:WindingBSM}
\eea

\noi Note that it is convenient to normalize the BSM by sui\-ta\-bly choosing the spacetime profile of $\bm{{\cal B}}(\bm{X})$ so that $Q_{\bm{{\cal B}}}=1$ or $w_{\bm{{\cal B}}}=1$. In the remainder, I solely consider spacetime textures which are characterized by a winding of a single positive unit of spacetime winding/vorticity.

This section concludes with a technical note regarding the construction of the line integral in Eq.~\eqref{eq:BSM} and its derivative definitions discussed in this section. Since in the present case there are two branches of Berry singularities in synthetic space, and not isolated points, it is required to avoid the contribution from other singularities along the direction of the line. This can be ensured by considering a di\-scre\-te implementation of the line integral, in which case, the integral $\int_0^{2\pi}d\vartheta/2\pi$ needs to be replaced by the sum $\nicefrac{1}{L}\sum_n$ where $n=1,\ldots,L$, with $L$ corresponding to the number of points com\-prising the discrete loop. Considering a discrete loop for the construction of the BSM is the most natural approach, since in experiments the spatial/time resolution of the probe cannot reach zero.

\section{Quantitative Analysis of the BSM in the ``clean'' case}\label{sec:SectionIX}

It is straightforward to numerically confirm these conclusions by evaluating the BSM defined in, for instance, Eq.~\eqref{eq:BSM} for a texture of the type $\bm{{\cal B}}(\vartheta)={\cal B}(\cos\vartheta,\sin\vartheta)$ with $w_{\bm{{\cal B}}}=1$. Here, $\vartheta\in[0,2\pi)$ parametrizes the synthetic texture loop. Within the semiclassical limit, the eva\-lua\-tion of the BSM relies on employing the bulk Hamiltonian which is assumed to adiabadically follow the spatially varying parameters $\bm{{\cal B}}(\vartheta)$. Therefore, within this limit, $\vartheta$ plays the role of either a spatial, or a time, or a spacetime coordinate. In this section, I numerically explore the BSM for the model of Eq.~\eqref{eq:gVector} with $M=0$ using the definition contained in the first line of Eq.~\eqref{eq:WindingBSM}.

\begin{figure}[t!]
\begin{center}
\includegraphics[width=0.95\columnwidth]{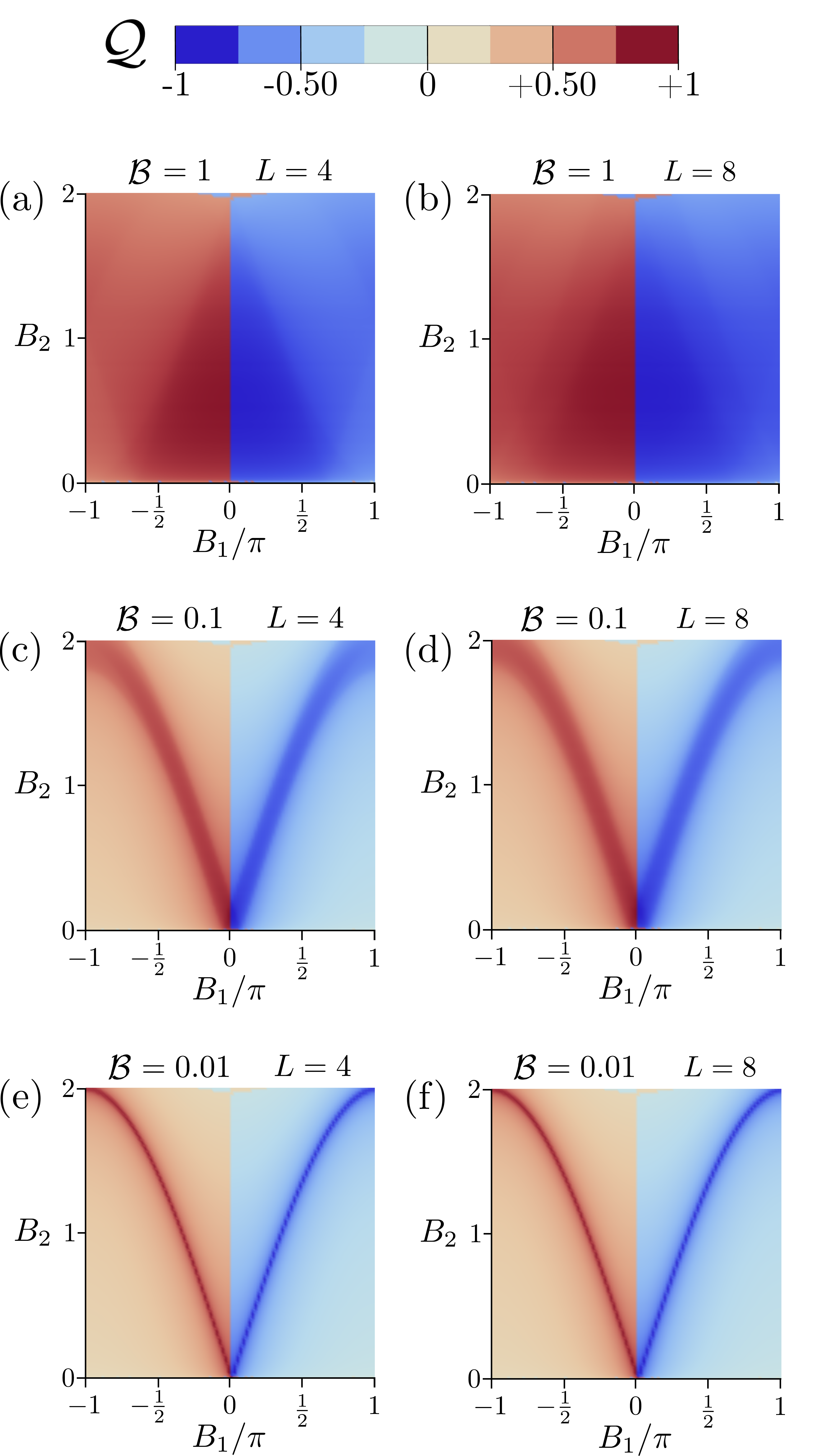}
\end{center}
\caption{Numerically evaluated Berry singularity marker (BSM). The panels depict a heat map of the BSM value ${\cal Q}$, according to the colorcoding presenting on the top of the fi\-gu\-re. The paramerer window coincides with the one considered in Figs.~\ref{fig:Figure1} and~\ref{fig:Figure2}. Panels \{(a),\,(c),\,(e)\} and \{(b),\,(d),\,(f)\} are calculated for a different period of the synthetic texture imposed on the spatiotemporally varying fields $\bm{{\cal B}}={\cal B}(\cos\vartheta,\sin\vartheta)$, as this is parametrized by the angle $\vartheta\in[0,2\pi)$. In the sequences \{(a),\,(c),\,(e)\} and \{(b),\,(d),\,(f)\}, the resolution increases from left to right, according to the three values ${\cal B}=\{1,\,0.1,\,0.01\}$. One concludes that enhan\-cing the resolution is crucial to be in a position to discern the Berry singu\-la\-ri\-ties through the desired quantization of the BSM. In contrast, increasing the period of the texture, i.e., the sampling points, does not bring drastic improvements to the BSM outcome. Therefore, \textit{in the clean case}, a discrete loop with a few points is sufficient to capture the presence of Berry singularities. In the calculations, ${\rm Tr}\big(\check{\chi}_{\rm BS}\big)$ was replaced by ${\rm Tr}\big[\check{\chi}\big(B_1=\cos^{-1}\big(B_2-1\big),B_2\big)\big]$, which is the respective value obtained for the nearest Berry singularity to point $\bm{B}$.}
\label{fig:Figure6}
\end{figure}

The respective numerical results are presented in Fig.~\ref{fig:Figure6}. In this set of results I have considered the three different values ${\cal B}=\{1,\,0.1,\,0.01\}$, which are ordered from left to right according to the increasing resolution of the experimental probe that these represent. In addition, I have assumed a synthetic texture of four and eight sites, i.e., $L=\{4,\,8\}$. The numerical results confirm two anti\-ci\-pa\-ted trends.  First of all, increasing the resolution by decreasing ${\cal B}$ can select out the Berry singularity branches. As seen from the respective panels, ${\cal B}$ essentially determines the width of a strip which encloses the singularity branches. This strip narrows down by reducing ${\cal B}$. The second expected feature inferred from these results concerns the ``sharpening'' of the BSM thanks to the increase of the number of sampling points $L$. Nonetheless, this improvement ends up to be very small here, thus, revealing that synthetic textures with short lengths $L$ can already capture the presence of the Berry singularities quite satisfactorily. As is discussed in the next section, this is not the case when disorder is introduced, in which case it is desirable to either increase $L$ or the number of cycles $N_c$, in order to apply the lock-in technique and improve the BSM outcome.

\begin{figure*}[t!]
\begin{center}
\includegraphics[width=1\textwidth]{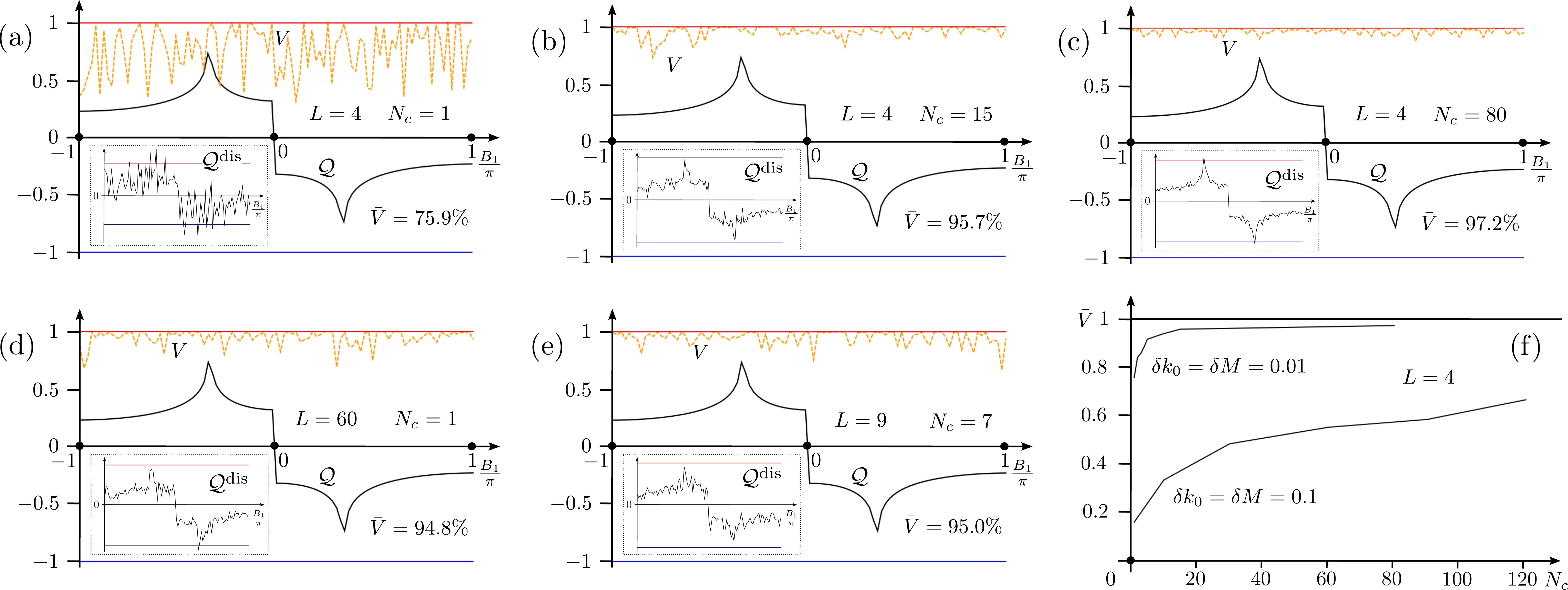}
\end{center}
\caption{Numerical investigation of the impact of disorder on the BSM along a cut of Fig.~\ref{fig:Figure6} for fixed $B_2=1.5$. Panels (a)-(c) show the obtained BSM in the absence of disorder (${\cal Q}$), the visibility (${\cal V}$), the averaged visibility ($\bar{{\cal V}}$) over the results obtained for the considered $B_1$ values, and the inset depicts the respective BSM in the presence of disorder (${\cal Q}^{\rm dis}$). In these panels the period of the synthetic texture is the same, but the number of sampling cycles vary. It becomes obvious that the visibility increases with the increase of $N_c$. Panels (d) and (e) contain similar results as the previous ones, but with the consideration of alternative strategies in regards to the length of the texture and the sampling cycles. It appears that for comparable number of the total sampling points $LN_c$, which are \{60,\,60,\,63\} in panels \{(b),\,(d),\,(e)\} of interest, maximizing the number of cycles yields the optimal result. The first five panels discuss random disorder of a strength $\delta M=0.01$. Panel (f) shows the averaged visibility obtained for fixed $L=4$, as a function of the number of sampling cycles $N_c$, for two different strengths of disorder. The plots were obtained by a connecting isolated points whose values are provided in App.~\ref{App:AppendixB}. One observes that increasing $N_c$ improves the visibility with, as expected, stronger disorder requiring more cycles to be suppressed. In all the above calculations I used ${\cal B}=0.001$ and considered a sampling step of $0.01$ in terms of $B_1/\pi$ to reduce the simulations runtime. In contrast, the calculations in Fig.~\ref{fig:Figure6} were performed for a step $0.001$ which captured the quantization of the peak, which is however missed here. Nonetheless, this does not influence the general conclusions. The red (blue) line corresponds to $1$ ($-1$) also in the inset.}
\label{fig:Figure7}
\end{figure*}

\section{Disorder And Lock-In Tomography}\label{sec:SectionX}

Up to this point, I have discussed and numerically eva\-lua\-ted the BSM in ``clean'' systems. However, several types of noise and disorder are generally present in the platforms studied in experiments. Therefore, in order to address such a realistic scenario, I here consider the pre\-sen\-ce of disorder, and demonstrate that the construction of BSM paves the way for disorder-free detection of Berry singularities and, in turn, topological phase transitions. 

In the present work, disorder is introduced in the model of Eq.~\eqref{eq:gVector} by considering that $M$ is spatiotemporally fluctuating. Within the adiabatic framework adopted here, it is not necessary to specify the origin and characteristics of these fluctuations. The key consequence of disorder here is that, when one eva\-lua\-tes the BSM by considering (for instance) a circular path parametrized by $\bm{{\cal B}}(\vartheta)={\cal B}(\cos\vartheta,\sin\vartheta)$, and $M$ is also dependent on $\vartheta$. In the remainder, I consider Anderson-type of disorder, that is, the pa\-ra\-me\-ter $M$ takes random values in the symmetric interval $[-\delta M,\delta M]$ with $\delta M\geq0$. Note that the variations in the system's parameters can be effectively absorbed by redefining $({\cal B}_1,{\cal B}_2)$, as $({\cal B}_1+\delta{\cal B}_1,{\cal B}_2+\delta{\cal B}_2)$.

It is instructive to first shed light on the effects of weak disorder, i.e., with $\delta {\cal B}_{1,2}$ smaller or comparable to the energy resolution of the experimental probe ${\cal B}$. In this case, the integral $I_s^{\rm dis}$ defined as in Eq.~\eqref{eq:Integral4BSM} but now in the presence of disorder, can be expanded up to linear order in both ${\cal B}_{1,2}$ and $\delta {\cal B}_{1,2}$, which in the spirit of Eq.~\eqref{eq:Integral4BSMapprox} here leads to the approximate expression:
\begin{align}
\frac{I_s^{\rm dis}-I_s}{I_s}\approx-\frac{1}{\pi}\ointctrclockwise_{{\cal C}_{\bm{B}_s}}\sum_{n,m}\frac{d\hat{{\cal B}}_n\varepsilon_{nm}\chi_{m\ell}(\bm{B}_s)}{{\rm Tr}\big[\check{\chi}(\bm{B}_s)\big]}\frac{\delta{\cal B}_\ell}{{\cal B}}\,,
\label{eq:Integral4BSMapproxDis}
\end{align}

\noi where $I_s$ defines the contribution for zero disorder and is given by Eq.~\eqref{eq:Integral4BSMapprox}. Besides immediately spoiling the possible quantization of the BSM, the presence of disorder can effectively reduce the resolution and drastically suppress the BSM near the Berry singularities, thus, leading to situation akin to the ones shown in Fig.~\ref{fig:Figure6}(a) and (b). 

In order to remedy these issues, I here propose to employ an experimental protocol which follows a principle which bears close si\-mi\-la\-ri\-ties to the one underlying the operation of the lock-in amplifier, cf Ref.~\onlinecite{LockIn}. This principle is a mere consequence of the orthogonality of sinusoidal functions. The potential applicability of this principle becomes apparent in the present situation after consi\-de\-ring a circular path for the loop ${\cal C}_{\bm{B}}$. For this case, the last term of Eq.~\eqref{eq:Integral4BSMapproxDis} retains contributions of the form: $\int_0^{2\pi}d\vartheta\{\cos\vartheta,\sin\vartheta\}\delta{\cal B}_{1,2}(\vartheta)$. When the variations $\delta{\cal B}_{1,2}(\vartheta)$ are either completely independent of $\vartheta$, or, if they can be expanded in a Fourier series consisting of terms containing higher harmonics of $\vartheta$, the above contributions drop out. In any other situation, disorder needs to be fully accounted for when calcu\-la\-ting the BSM. This problem constitutes a typical issue in signal processing when noise is present. There, the contribution of disorder becomes decoupled by considering a sufficiently large number $N_c$ of sampling cycles, so that the integrals ave\-ra\-ge out to zero.

To demonstrate how this lock-in tomography of Berry singularities becomes realized, I first consider the case of weak disorder. I first focus on this regime to infer the optimal strategies that lead to suppression of the contributions of di\-sorder to the BSM. To quantify the impact of disorder on the BSM, I introduce the visibility~\cite{CommentVis}:
\begin{align}
{\cal V}=1/\sqrt{1+\big(1-{\cal Q}^{\rm dis}/{\cal Q}\big)^2}
\end{align}

\noi which takes values in the range $[0,1]$. The visibility becomes equal to $1$ ($0$) when the value of ${\cal Q}^{\rm dis}$ is (not) equal to its counterpart ${\cal Q}$ defined in the absence of di\-sorder.

To carry out the evaluation of ${\cal V}$, I consider once again a discrete circular path in the synthetic space, which is parametrized by $\vartheta=2\pi n/L$, with $n=1,2,3,\ldots,L$, with $L$ denoting the number of sites of the synthetic texture. With no loss of generality, I restrict to a single cut of the synthetic space, with $B_2=1.5$. Similar general conclusions can be drawn for values for $B_2$. The disorder profile is here generated using a standard random ge\-ne\-ra\-tor, and every result for a given $B_1$ is averaged over twenty random disorder rea\-li\-za\-tions.  

In the first set of numerical results, I study how the vi\-si\-bi\-li\-ty becomes enhanced by increasing $N_c$. Specifically, I fix the synthetic texture length to the value $L=4$ and vary the number of sampling cycles $N_c$, i.e., the integration variable $\vartheta$ in Eq.~\eqref{eq:WindingBSM} is now defined in the interval $[0,2\pi N_c)$. Indeed, from panels (a)-(c) in Fig.~\ref{fig:Figure7}, one confirms the relatively fast restoration of the visibility since we are in the limit of weak disorder $\delta M=0.01$. One also notes, however, that there exist spikes in the dependence on $B_1$, and a generally not so smooth evolution. This is a result of the random disorder considered here. Hence, one should mainly focus on the general trend rather than exact evolution of the profile. In ano\-ther batch of results, which are presented in Fig.~\ref{fig:Figure7}(f) I plot the ave\-ra\-ged visibility obtained for $L=4$ and various values of $N_c$. At this point it is natural to ask which approach is preferred for the same number of total sampling points. That is, assuming that the product $LN_c$ is fixed, what is the optimal ratio between $L$ and $N_c$ which suppresses noise faster, i.e., for the smallest possible product $LN_c$. An exploration of this sort is presented in panels (d) and (e) of Fig.~\ref{fig:Figure7}. By comparing the results of panels (b), (d) and (e) it emerges that the optimal strategy is to ma\-xi\-mi\-ze $N_c$. Nevertheless, further investigation is required to reach to generic conclusions.

There is no obvious reason not to expect that the ge\-ne\-ral trends obtained in the weak disorder limit discussed above, do not carry over to the moderate or strong disor\-der case, in which situation the approximate form in Eq.~\eqref{eq:Integral4BSMapproxDis} does not apply any longer. In panel Fig.~\ref{fig:Figure7}(e) I also present results which address the case of stronger disorder with $\delta M=0.1$. Indeed, it becomes clear that similar trends hold also in the present scenario, but the number of cycles $N_c$ required to suppress the effects of disorder is, without surprise, significantly larger. Additional details regarding the calculations in this limit are provided in App.~\ref{App:AppendixB}. There, I provide further details on the precise results and methods I used to obtain Fig.~\ref{fig:Figure7}(e).

\section{Applications of BSM approach}

\begin{figure}[t!]
\begin{center}
\includegraphics[width=\columnwidth]{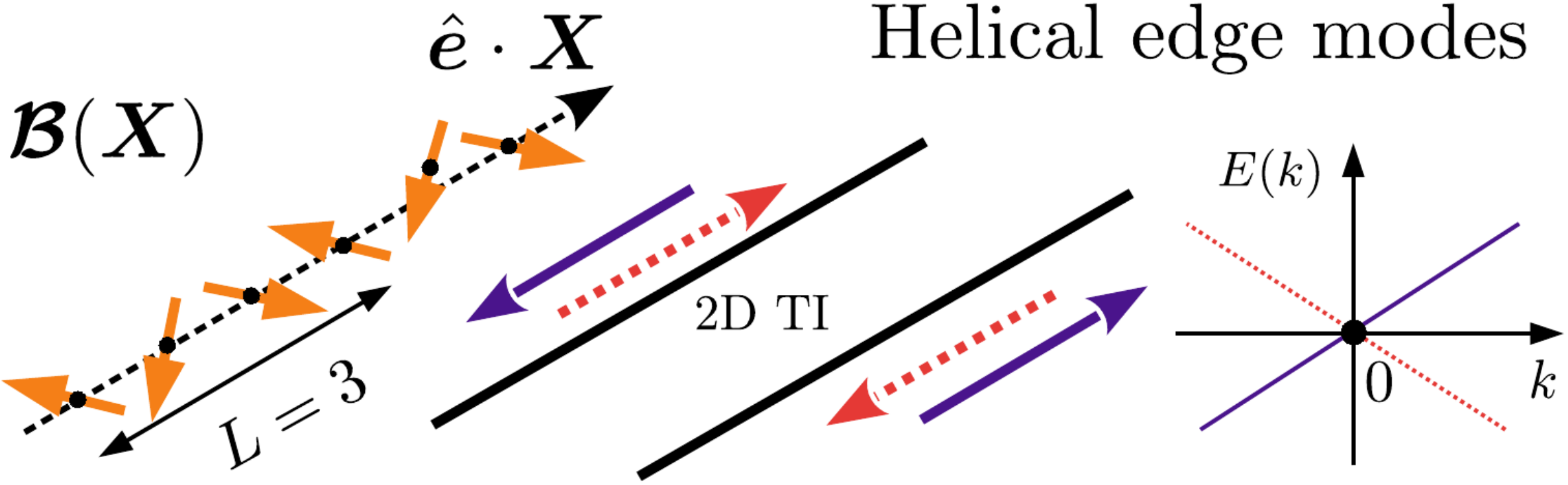}
\end{center}
\caption{Experimental setup for benchmarking the theory of Berry singularity markers proposed here. The helical modes on a given edge of a 2D topological insulator can be viewed as a robust Berry singularity, which is protected by time-reversal symmetry. I propose to verify the topological nature of the crossing by evaluating the BSM in the presence of a spiralling magnetic field. Carrying out this experiment in a known 2D topological insulator, such as HgTe quantum wells~\cite{HgTe}, in the additional presence of intrinsic or controllable disorder, can provide a testbed for the implementation of BSMs.}
\label{fig:Figure8}
\end{figure}

The BSM method that I have brought forward in this work appears directly testable in a variety of currrently-accessible expe\-ri\-mental platforms. The concrete application that I discuss here concerns employing the BSM approach to identify the topologically-protected crossing of the helical modes on a given edge of a 2D topological insulator (TI). When $k_0=M=0$ and $\tau$ corresponds to spin, a suitably linearized version of the model in Eq.~\eqref{eq:gVector} describes the massless helical edge modes which appear on a given edge of a 2D topological insulator~\cite{HasanKane,QiZhang}, such as HgTe quantum wells~\cite{HgTe}. See Fig.~\ref{fig:Figure8} for a schematic.

Benchmarking the BSM approach in such a system requires to experimentally infer the BSM by studying the magnetic response of the helical edge modes to a Zeeman or exchange field $\bm{B}+\bm{{\cal B}}(\bm{X})$, which consists of a constant offset $\bm{B}$, and a component $\bm{{\cal B}}$ which gives rise to a spacetime texture. In the minimal realization, the latter texture can be for instance a magnetic spiral which unfolds in the direction of spacetime $\bm{X}$ defined by the unit vector $\hat{\bm{e}}$, as depicted in Fig.~\ref{fig:Figure8}. The spiralling part of the magnetic field is required to wind in a spin plane which involves the direction determined by the polarization of the helical edge modes. Hence, the adiabatic $\bm{d}$ vector which describes the si\-tua\-tion of interest reads as:
\begin{align}
\bm{d}(\bm{K},\bm{B}+\bm{{\cal B}}(\bm{X}))=\big(k-B_1-{\cal B}_1(\bm{X}),-B_2-{\cal B}_2(\bm{X}),0\big).
\end{align}

\noi The above gives rise to a Hamiltonian which bear si\-mi\-la\-ri\-ties to the one discussed in Eq.~\ref{eq:gVector}, with the dif\-fe\-ren\-ce that here one is restricted to the vicinity of $k=0$, and the value for the parameter $M$ is equal to zero (assuming $k_0=0$). The latter property also manifests the key difference between the two systems. Specifically, the here-considered system is exactly tuned to a ``Berry sin\-gu\-la\-ri\-ty'', i.e., $M=0$. To be precise, what plays the role of a Berry singularity here is actually the topologically-protected touching point of the two spin-filtered counter-propagating modes~\cite{KaneMele1,KaneMele2}, rather than a true Berry singularity dictating the gap closings of a bulk band structure.

In spite of the important conceptual differences, the Hamil\-to\-nians describing the two situations present strong analogies which allows us to apply the already extracted BSM results also to the TI edge case. As a matter of fact, the entire prior analysis is directly applicable to the TI edge system, as long as (i) we restrict to va\-lues of $|\bm{B}|$ which are much smaller than the bulk gap of the 2D TI, and (ii) identify the conjugate vector field $\bm{P}$ and susceptibility $\check{\chi}$, with the magnetization field $\bm{{\cal M}}=-\partial E_{\rm edge}/\partial \bm{B}$ and magnetic susceptibility $\chi_{\rm edge}^{nm}=\partial {\cal M}_n/\partial B_m$ of the TI edge. Here, $E_{\rm edge}$ denotes the energy contribution of the TI edge at zero temperature. Finally, it is important to remark that here the diamagnetic response is negligible and the chiral susceptibility is obtained from the response to a magnetic/exhange field coupling to the helical edge modes through the Hamiltonian term $\propto B_3\tau_3$.

\section{Conclusions and Outlook}\label{sec:SectionXI}

In this work, I introduce an alternative method to experimentally characterize topological phase transitions in insulators, which relies on detecting the Berry singularities of the system. The latter are defined in a synthetic space $\bm{K}=(\bm{k},M)$, spanned by the wave vector $\bm{k}$ labelling the Bloch bands, and a parameter $M$ which controls the bulk energy gap. Here, I introduce a quantity coined Berry singularity marker (BSM), which signals the detection of a Berry singularity. To evaluate the BSM, one relies on studying the response of the sy\-stem in the presence of a set of suitable external fields $\bm{F}$, which ena\-ble variations in all components of $\bm{K}$, and perform a mapping of the Berry singularities to the synthetic space they span. As I show, the BSM becomes sizeable only in the vicinity of Berry singularities, and its value is mainly determined by the susceptibility tensor elements $\chi_{11,22,12}(\bm{B})$. The off-diagonal susceptibility element determines the sign of the Berry singularity charge, while the trace of the diagonal sucsceptibility elements provide a notion of distance from the Berry singularity and is in fact related to the trace of the quantum metric~\cite{Provost}.

Although previous works~\cite{CPSun,Zanardi,HQLin,Ma1,Matsuura,Ma2,Montambaux} have highlighted the value of the quantum metric to detect phase transitions, these had not focused on the information which becomes accessed with respect to the Berry singularities. The construction of the BSM that I present here provides a phy\-sical quantity which behaves similar to the Berry charge density and, in fact, tends to the topological charge of the Berry singularity it detects. Even more, the BSM further accounts for the sign of the Berry singularity which also yields information regarding the sign of the variation of the topological invariant characterizing the bulk system, and not only its critical line. However, the most re\-mar\-ka\-ble feature of the BSM, is that it can be implemented using spacetime textures of the external fields, which are in this case decomposed as $\bm{F}=\bm{B}+\bm{{\cal B}}(\bm{X})$. The constant offsets $\bm{B}$ allow identifying the locations of the Berry singularities in the original $\bm{K}$ space, while the varying part $\bm{{\cal B}}(\bm{X})$ encodes the spacetime synthetic texture. These, in turn, enable the lock-in evaluation of the BSM, therefore providing a unique path towards the noise-free experimental detection of topological phase transitions.

To demonstrate the above properties, in this work I employ a representative AIII topological insulator model, which shares similarities to the celebrated Su-Schrieffer-Heeger (SSH) model~\cite{SSH}. Besides the explicit construction of the BSM in this case, I further carry out a nu\-me\-ri\-cal investigation of the BSM in the absence and presence of disorder. In the former case, I study the behaviour of the BSM when various parameters are changed, such as the resolution of the experimental probe, which is represented by the amplitude of the synthetic texture $\bm{{\cal B}}(\bm{X})$. In connection to disorder, I consider various rea\-li\-za\-tions of weak and moderate strengths of random di\-sor\-der which effectively modify $\bm{{\cal B}}(\bm{X})$. My numerical investigations provided evidence for the success of the approach in suppressing the influence of noise/disorder on the evaluation of the BSM. Similar to the principle of the lock-in amplifier, also here the full suppression of disorder is in principle possible by accordingly increasing the number of sampling points. Despite the fact that in a real experiment this approach may face limitations, the present method yet brings forward the BSM as a quantity which encodes all the information regarding the to\-po\-lo\-gi\-cal phase transition, and can be rendered immune to noise and disorder in a straightforward and generic way.

The analysis which I present in this work appears directly testable in two types of currrently-accessible expe\-ri\-ments. First of all, when $k_0=M=0$ and $\tau$ corresponds to spin, the model in Eq.~\eqref{eq:gVector} effectively describes the massless helical edge modes which appear on a given edge of a 2D topological insulator~\cite{HasanKane,QiZhang}, such as HgTe quantum wells~\cite{HgTe}. See Fig.~\ref{fig:Figure8} for a schematic. Notably, this Hamiltonian correponds to a situation where the system is exactly tuned to a Berry singularity, i.e., which corresponds to the bulk-topology protected crossing point of the two spin-filtered counter-propagating modes~\cite{KaneMele1,KaneMele2}. Therefore, to benchmark the present method, one can experimentally infer the BSM by studying the magnetic response of the system to a Zeeman or exchange field which forms a spacetime texture. The spiralling magnetic field is required to wind in a spin plane which involves the direction determined by the polarization of the helical edge modes.

One can envisage natural extensions in higher dimensions, where one can in principle follow the very same thinking and construct BSMs based on the response functions of the systems $\bm{P}(\bm{X})$. From Eq.~\eqref{eq:WindingBSM} one observes that the BSM can be re-written in terms of derivatives of the conjugate fields, which is closer to the usual construction of the winding number using the $\bm{d}$-vector of the Hamiltonian. Therefore, suitable antisymmetrized derivatives of $\bm{P}$ with respect to $\bm{{\cal B}}(\bm{X})$ appear promising to lead the construction of the BSM in higher-dimensional systems where the Bloch functions support a winding or a Chern number. Special attention has to be paid, however, when it comes to superconducting sy\-stems in classes BDI, D, DIII, since the charge-conjugation symmetry present does not allow to mix functions of a different parity as it was possible here. Elsewhere, I propose BSMs which circumvent this obstacle~\cite{PKtoappear}.

\begin{figure}[t!]
\begin{center}
\includegraphics[width=0.8\columnwidth]{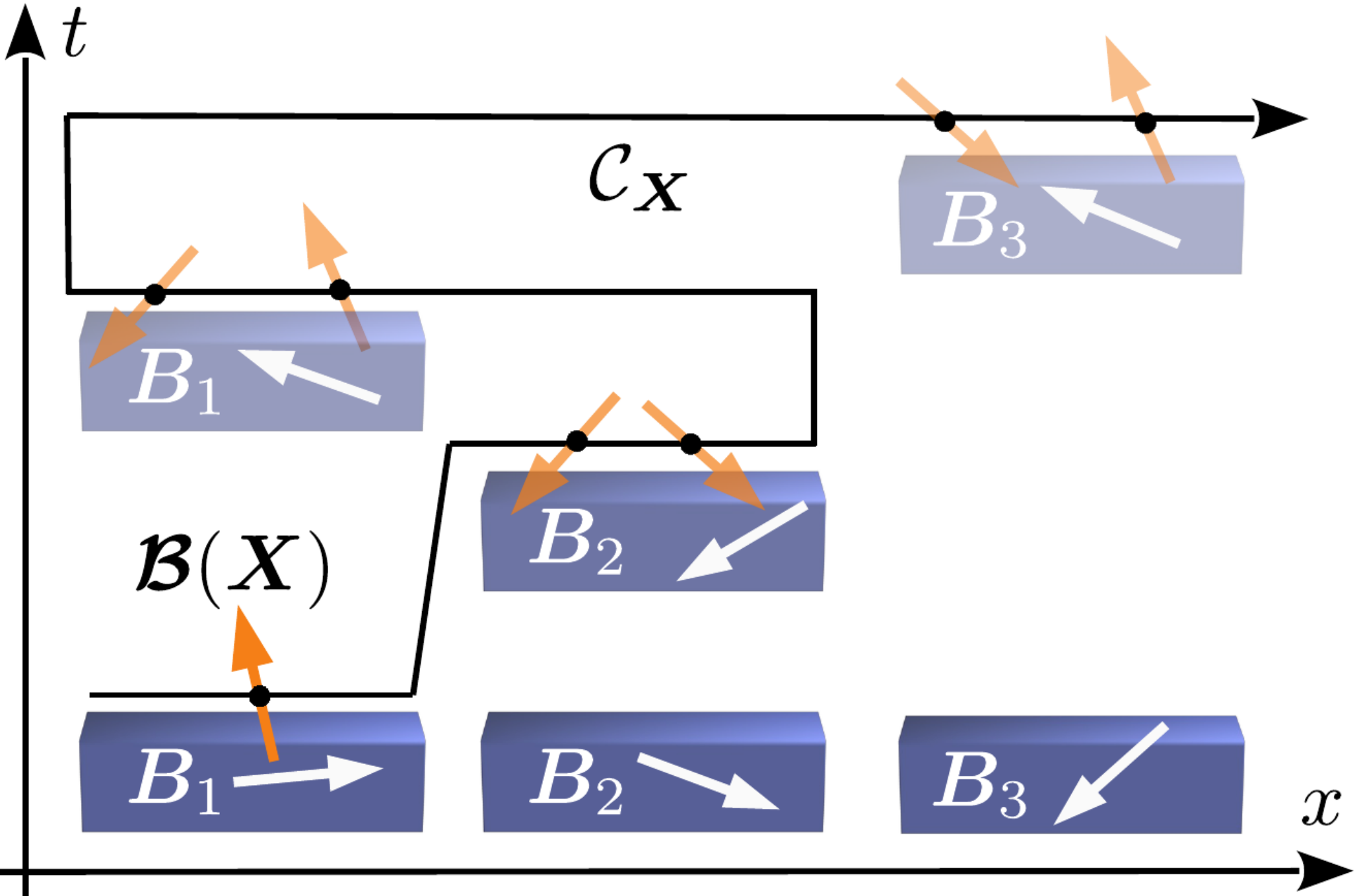}
\end{center}
\caption{Sketch for a spacetime texture constructed using multiple systems which are in principle identical. The degree of topological similarity of the ensemble can be quantified by constructing the ensemble BSM. Possible detunings among the parameter values of the systems of the ensemble function as disorder which can be imprinted in the outcome of the ensemble BSM. Hence, the ensemble BSM in this case provides a measure of collective topological coherence, which can be useful for scaling up devices targetted for quantum compu\-ting and other applications. Finally, note that I only placed the three systems along a single axis for simpilicity. In ge\-ne\-ral, ${\cal C}_{\bm{X}}$ can be any closed loop in quasi-3D or 4D spacetime.}
\label{fig:Figure9}
\end{figure}

Concluding this work, I remark that the construction of the BSM implies that its applicability goes beyond the characterization of only a single system. Indeed the spatial coordinates of $\bm{X}$ which enter in the evaluation of the BSM may correpond to the locations of completely disconnected systems which are expected to be described by the same Hamiltonian. See for instance Fig.~\ref{fig:Figure9} for the depiction of such a spacetime loop. One expects that when the $\bm{B}$ components are tuned to be equal for all the systems involved, and in the absence of disorder, the result of the ensemble BSM will provide a measure of the collective to\-po\-lo\-gi\-cal coherence. Such a measure appears particularly useful in the process of scaling-up topo\-lo\-gi\-cal systems in order to use them for applications such as universal quantum computing. Therefore, employing the BSM using spacetime textures involving the various components of the device can provide a measure for the topological functionality of the ensemble.  

\section*{Acknowledgements}

I acknowledge funding from the National Na\-tu\-ral Science Foundation of China (Grant No.~12074392 and No.~12050410262).

\appendix

\section{Alternative coupling to the fields}\label{App:AppendixA}

The presentation and numerical ana\-lysis of the main text have been both based on the specific type of coupling of the system to the external field discussed in Sec.~\ref{sec:SectionIII}. However, as I already remarked in Sec.~\ref{sec:SectionIII}, the choice of couplings is not unique, and this varies depending on the system of interest. It is in fact possible that in several experimental situations enginee\-ring the coupling considered in Sec.~\ref{sec:SectionIII} may be proven hard to achieve. To tackle such potential shortcomings, in this paragraph I investigate an alternative possibility. Specifically, I consider the situation where the external fields enter the Hamiltonian by shifting the isospin vector $\bm{d}(\bm{K})$, i.e., $\bm{d}(\bm{K})\mapsto\bm{d}(\bm{K})-\bm{F}(\bm{X})$. In this case, the ener\-gy dispersions of the system become modified according to $E_\pm[\bm{K},\bm{F}(\bm{X})]=\pm|\bm{d}(\bm{K})-\bm{F}(\bm{X})|$. 

\begin{figure}[t!]
\begin{center}
\includegraphics[width=1\columnwidth]{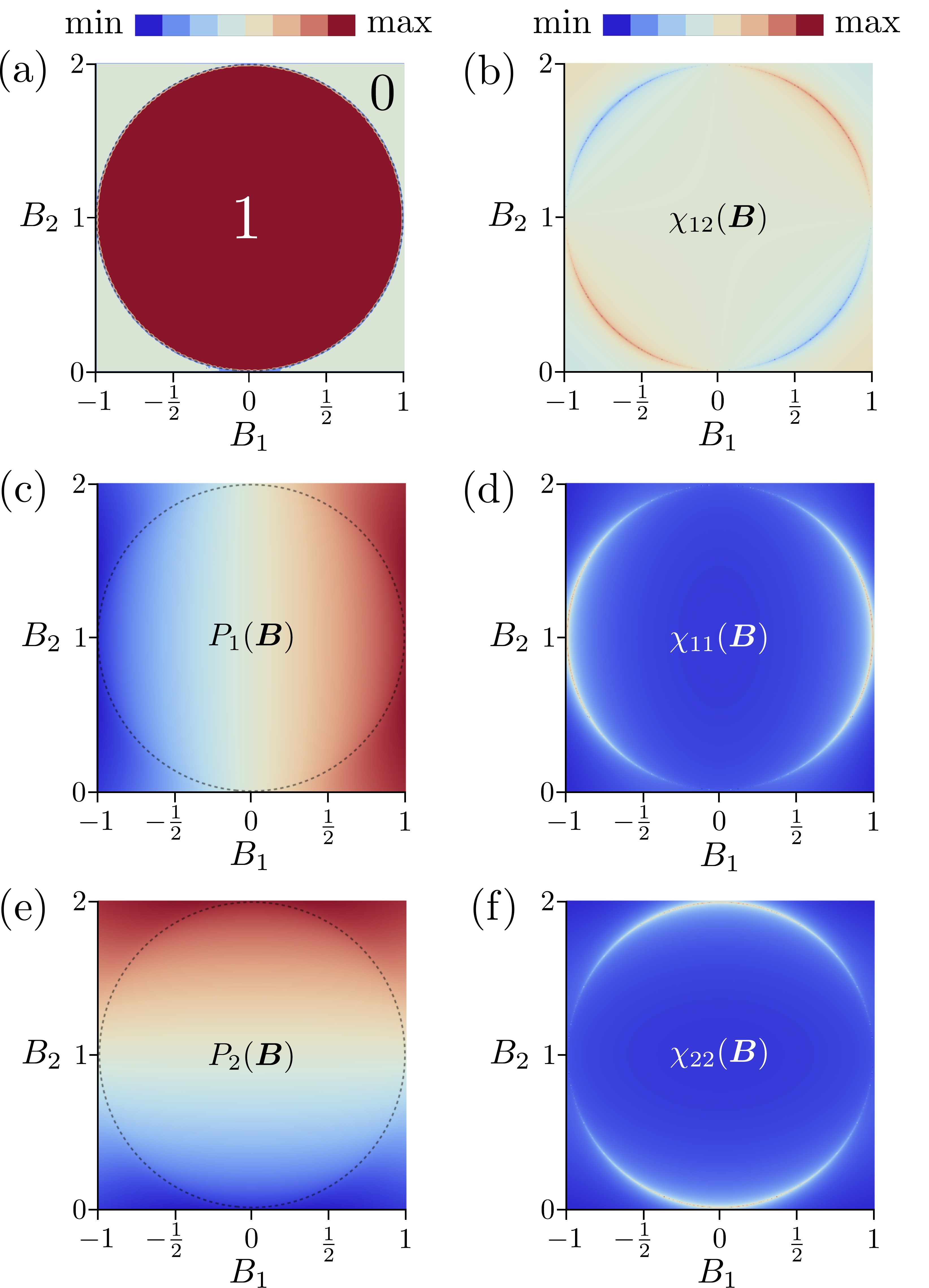}
\end{center}
\caption{Properties of the AIII topological insulator with the alternative coupling $\bm{d}(\bm{K})\mapsto\bm{d}(\bm{K})-\bm{F}(\bm{X})$ to the external fields $\bm{F}(\bm{X})$. (a) Topological phase diagram and Berry singularity branches (dot-dashed lines) defined in the synthetic space $\bm{B}$. Panels (b) and (c) depict the numerically evaluated conjugate fields $\bm{P}(\bm{B})$, while panels (d)-(f) depict the numerically evaluated susceptibility tensor elements $\chi_{11,22,12}(\bm{B})$. Besides the lack of periodicity in terms of $B_1$, the obtained results are very similar to the ones of Figs.~\ref{fig:Figure2} and~\ref{fig:Figure3}.}
\label{fig:Figure10}
\end{figure}

In order to infer the fate of the BSM in the present case, I now proceed by examining the modifications introduced to the Berry singularity branches in synthetic space, as well as to the response of the system, due to the updated coupling to the external fields. With the new coupling, and ${\cal B}=0$, the Berry singularities are now given by simultaneously satisfying the gap-closing conditions $\sin k=B_1\quad{\rm and}\quad 1-\cos k+M=B_2$. As a consequence, the positions of the Berry singularities in the synthetic $\bm{B}$ space are given by the geometric locus:
\bea
B_1^2+\big(B_2-M-1\big)^2=1
\eea

\noi which corresponds to a circle centered at the synthetic space point $\bm{B}=(0,M+1)$. The locus of Berry singularities is shown in Fig.~\ref{fig:Figure10}(a), and is superimposed on a heat map depicting the topological invariant of the system obtained as a function of $(B_1,B_2)$. Direct calculation of the winding number for $M=0$, yields that the topologically-nontrivial phase with $w=1$ appears at the disc enclosed by the Berry singularity circle. 

As a next step, I investigate the response of the sy\-stem to the external fields. Once again, by considering the zero-temperature limit, one finds that the conjugate vector field $\bm{P}[\bm{F}(\bm{X})]$ now reads as:
\bea
\bm{P}[\bm{F}(\bm{X})]=-\int_{\rm BZ}\frac{dk}{2\pi}\ph\frac{\bm{d}(\bm{K})-\bm{F}(\bm{X})}{|\bm{d}(\bm{K})-\bm{F}(\bm{X})|}\,.
\eea

\noi The conjugate fields as functions of $(B_1,B_2)$ are shown in panels (c) and (e) of Fig.~\ref{fig:Figure10}, and bear similarities to their counterparts in panels (a) and (c) of Fig.~\ref{fig:Figure2}, which are obtained for the initial type of coupling. One confirms that the corresponding conjugate field components share the same type of symmetries under mirror operations about the high-symmetry lines. However, they still present a difference, that is, in the present coupling the conjugate fields do not present any periodicity with respect to $B_1$.

\begin{figure*}[t!]
\begin{center}
\includegraphics[width=1\textwidth]{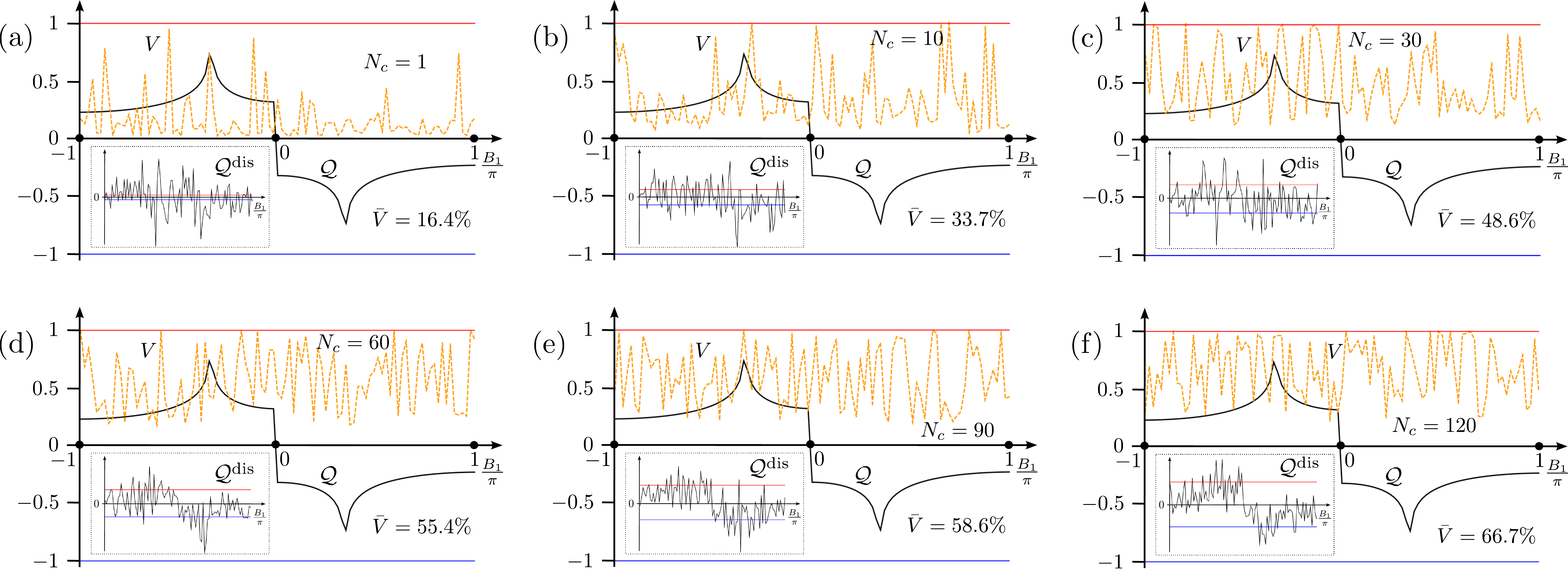}
\end{center}
\caption{Numerical investigation of the impact of disorder on the BSM along a cut of Fig.~\ref{fig:Figure6} for fixed $B_2=1.5$ and $\delta M=0.1$. Panels (a)-(f) show the obtained BSM in the absence of disorder (${\cal Q}$), the visibility (${\cal V}$), the averaged visibility ($\bar{{\cal V}}$) over the results obtained for various $B_1$ values, and the inset depicts the respective BSM in the presence of disorder (${\cal Q}^{\rm dis}$). In these panels the period of the synthetic texture is the same, but the number of sampling cycles vary. It becomes obvious that the visibility increases with the increase of $N_c$ but this requires more cycles compared to the case of weaker disorder as confirmed by Fig.~\ref{fig:Figure7}(e). In all the above calculations I used ${\cal B}=0.001$ and considered a sampling step of $0.01$ in terms of $B_1/\pi$ to reduce the simulations runtime. Finally, note that the red (blue) line corresponds to $1$ ($-1$) also in the inset.}
\label{fig:Figure11}
\end{figure*}

The susceptibilities are also obtained in a fashion ana\-lo\-gous to the one discussed in Sec.~\ref{sec:SectionV}. The related outcomes are shown in panels (b), (d) and (f) of Fig.~\ref{fig:Figure10}. Lea\-ving aside the lack of periodicity in terms of $B_1$, the here-found susceptibilities present a similar symmetry behaviour with their counterparts discussed in Sec.~\ref{sec:SectionV}. Most importantly, however, in the present case the diamagnetic contribution to the susceptibility is exactly zero, thus implying that the susceptibility is solely go\-ver\-ned by the quantum metric. 

Based on the above results, one infers that the construction and behaviour of the BSM for the present type of coupling is expected to be qualitatively similar to the one extensively discussed in this manuscript. For this reason I do not investigate further details of this model.

\section{Additional results regarding disorder}\label{App:AppendixB}

In this appendix, I provide further details regarding the way I constructed Fig.~\ref{fig:Figure7}(e). For a disorder strength equal to $\delta M=0.01$, I performed numerical investigations which provided the following results for $(N_c,\bar{V}(\%))$: $(1,75.9\%)$, $(2,84.1\%)$, $(3,85.7\%)$, $(4,88.4\%)$, $(5,91.6\%)$, $(10,93.9\%)$, $(15,95.7\%)$, and $(80,97.2\%)$. The resul\-ting plot was obtained by simply connecting the above points. A similar approach was followed for the plot in the stronger disorder case, using the results in Fig.~\ref{fig:Figure11}.


\begin{thebibliography}{99}
\bibitem{CommentBDI} Setting $k_0=0$ gives rise to a charge-conjugation symmetry, which is generated by the complex conjugation operator ${\cal K}$. This additional symmetry modifies the symmetry class according to AIII$\rightarrow$BDI. Nevertheless, this symmetry enhancement is insignificant for the results obtained in this work since, for topological insulators in one spatial dimension, the properties of the BDI class are directly inherited from those of the AIII class.
\bibitem{CommentCoupling}Note that the chosen prescription for the coupling is not unique. In fact, App.~\ref{App:AppendixA} discusses an alternative coupling. Furthermore, extra caution should be paid on the choice of the coupling and the nature of the fields $\bm{F}$ when discussing spatial textures for $\bm{F}(\bm{X})$. Specifically, one should avoid employing U(1) gauge fields for $\bm{F}$ in this case. This is because the gauge freedom allows to remove static U(1) gauge fields from the Hamiltonian and absorb them into twisted boundary conditions. However, such ``large'' gauge transformations are generally not equivalent to the spatial textures that are assumed throughout this work.
\bibitem{CommentHigher} For systems belonging to symmetry classes which allow for Berry charges higher than $\pm1$, one can infer the desired windings by means of higher-order derivatives of $\bm{P}(\bm{B})$ in terms of $\bm{B}$.
\bibitem{CommentUniversal} To be precise, the value of ${\rm Tr}\big(\check{\chi}_{\rm BS}\big)$ varies depending on the sy\-stem under examination. Nevertheless, it is universal in the sense that, for a given material, it takes distinct values which solely depend on the value of the topological charge of the Berry singularity.
\bibitem{CommentSection} As I discuss in Sec.~\ref{sec:SectionVII}, when the diamagnetic response is negligible, this ``universal'' constant can be alternatively obtained by experimentally inferring the expectation value of the chiral symmetry operator $\Pi$.
\bibitem{CommentStiffness}Notably the study of quantum geometry in static response has recently picked up si\-gni\-fi\-cant attention in relation to the properties of superfluid density of multiband superconductors harboring flat bands~\cite{Torma,LongLiang}. In the present work, the aim is clearly different, and concerns developing an approach to detect Berry singularities in a disorder-immune fashion.
\bibitem{CommentVis} Other choices for the visibility are not expected to alter the ge\-ne\-ral trends of the results.
\end{thebibliography}
\end{document}